\documentclass[preprint]{aastex63}

\usepackage{gensymb}
\turnoffeditone
\turnoffeditthree
\accepted{October 15, 2020}
\submitjournal{ApJ}

\shorttitle{VLA observations}
\shortauthors{Maithil et al.}
\graphicspath{{./}{figures/}}
\begin{document}

%\title{VLA Observations of a Sample of Radio-Loud Quasars Selected to Investigate Orientation Effects}

\title{Investigating orientation effects considering angular resolution for a sample of radio-loud quasars using VLA observations}

\correspondingauthor{Jaya Maithil}
\email{jmaithil@uwyo.edu}

\author[0000-0002-4423-4584]{Jaya Maithil}
\affil{Department of Physics and Astronomy, 
University of Wyoming, 
Laramie, WY 82071, USA }

\author{Jessie C. Runnoe}
\affiliation{Physics \& Astronomy Department,
Vanderbilt University, Nashville, TN 37235, USA}

\author{Michael S. Brotherton}
\affil{Department of Physics and Astronomy, University of Wyoming, Laramie, WY 82071, USA }

\author{John F. Wardle}
\affil{ Department of Physics, Brandeis University, Waltham, MA 02454, USA.}

\author{Beverley J. Wills}
\affil{ College of Natural Sciences, University of Texas at Austin, Austin, TX 78712, USA.}

\author{Michael DiPompeo}
\affil{Department of Physics and Astronomy, Dartmouth College, Hanover, NH 03755, USA.}

\author{Carlos De Breuck}
\affil{ European Southern Observatory, Garching, D-85748, Germany.}

%\collaboration{1}{(LaTeX collaboration)}

%\nocollaboration{2}

%% Note that the \and command from previous versions of AASTeX is now
%% depreciated in this version as it is no longer necessary. AASTeX 
%% automatically takes care of all commas and "and"s between authors names.

%% AASTeX 6.3 has the new \collaboration and \nocollaboration commands to
%% provide the collaboration status of a group of authors. These commands 
%% can be used either before or after the list of corresponding authors. The
%% argument for \collaboration is the collaboration identifier. Authors are
%% encouraged to surround collaboration identifiers with ()s. The 
%% \nocollaboration command takes no argument and exists to indicate that
%% the nearby authors are not part of surrounding collaborations.

%% Mark off the abstract in the ``abstract'' environment. 
\begin{abstract}
%\deleted{We have obtained continuum observations of 147 radio-loud quasars using the NSF's Karl G. Jansky Very Large Array (VLA) with the goal of determining orientation indicators, such as different measures of the radio core dominance. We used the VLA at 10 GHz in A-array to best spatially resolve the cores from extended emission. Orientation studies based on low-resolution radio surveys, like Faint Images of the Radio Sky at Twenty cm (FIRST), often fail to fully resolve cores. Our measurements show that at the FIRST spatial resolution of 5$\arcsec$, core flux measurements are indeed systematically high compared to our new 0.2$\arcsec$ resolution data. Our results empirically confirm that determination of radio core dominance requires high-spatial resolution data. Our result also shows that this effect is more prominent for quasars with smaller angular size.}

\edit1{Radio core dominance measurements, an indicator of jet orientation, sometimes rely on core flux density measurements from large-area surveys like Faint Images of the Radio Sky at Twenty cm (FIRST) that have an angular resolution of only 5$\arcsec$. Such low-resolution surveys often fail to resolve cores from the extended emission resulting in an erroneous measurement. We focus on investigating this resolution effect for a sample of 119 radio-loud quasars. We obtained continuum observations from NSF's Karl G. Jansky Very Large Array (VLA) at 10 GHz in A-array with a 0.2 $\arcsec$ resolution. Our measurements show that at FIRST spatial resolution, core flux measurements are indeed systematically high even after considering the core-variability. For a handful of quasars, 10 GHz images reveal extended features, whereas the FIRST image shows a point source. We found that the resolution effect is more prominent for quasars with smaller angular sizes. We further computed two radio core dominance parameters R \& $\rm R_{5100}$ for use in statistical orientation investigations with this sample. We also present the spectral energy distributions between 74 MHz and 1.4 GHz, which we used to measure the spectral index of the extended emission of these quasars. Our results empirically confirm that determination of radio core dominance requires high-spatial resolution data. We highlight the practical issues associated with the choice of frequency and resolution in the measurement of core and extended flux densities.} 

\end{abstract}

%% Keywords should appear after the \end{abstract} command. 
%% See the online documentation for the full list of available subject
%% keywords and the rules for their use.
%\keywords{editorials, notices --- 
%miscellaneous --- catalogs --- surveys}

%% From the front matter, we move on to the body of the paper.
%% Sections are demarcated by \section and \subsection, respectively.
%% Observe the use of the LaTeX \label
%% command after the \subsection to give a symbolic KEY to the
%% subsection for cross-referencing in a \ref command.
%% You can use LaTeX's \ref and \label commands to keep track of
%% cross-references to sections, equations, tables, and figures.
%% That way, if you change the order of any elements, LaTeX will
%% automatically renumber them.
%%
%% We recommend that authors also use the natbib \citep
%% and \citet commands to identify citations.  The citations are
%% tied to the reference list via symbolic KEYs. The KEY corresponds
%% to the KEY in the \bibitem in the reference list below. 

\section{Introduction} \label{sec:intro}

Quasars are the most luminous form of active galactic nuclei (AGNs), powered by the release of gravitational potential energy of material accreting onto the supermassive black holes at their centers. The axisymmetric nature of quasars makes orientation to the line of sight an essential quantity for comprehending their observed properties. Orientation plays a pivotal role in unifying not just type 1 and type 2 AGNs, e.g., Seyfert 1 and Seyfert 2 galaxies \citep{AM85}, but it is also fundamental in unifying classes of radio-loud quasars. When the radio jet axis of a quasar is aligned close to the line of sight \edit1{(LOS)}, the core radiation gets Doppler-boosted and we observe a flat-spectrum radio-loud quasar, whereas, when the \edit1{quasar is observed at larger angles to the LOS,} little Doppler-boosting takes place and we observe a steep-spectrum radio-loud quasar \citep{OB82} dominated by the optically thin synchrotron radiation of the \edit1{extended} radio lobes. Properties of the optical spectra differ between these classes.  The full width at half maximum (FWHM) of broad emission lines, such as  $\mbox H\beta$, correlates with orientation: \edit1{quasars aligned close to the LOS show lower values of FWHM whereas the quasars at larger angles to the LOS ($\sim 60^{\circ}$ in type 1 AGNs, \citealt{Marin}) show systematically higher values \citep{WB86}.}

Orientation not only governs classifications but also impacts the quantitative determination of fundamental properties like black hole mass and luminosity. The optical H$\beta$ line emitted from what is likely a flattened broad-line region is most commonly used to derive the mass of the black hole. As FWHM H$\beta$ correlates with the orientation \citep{WB86}, the derived mass is also orientation dependent (e.g., \citealt{Runnoe2013}). Bolometric luminosity determination also depends on orientation \citep{NB2010}. Face-on radio-loud quasars have brighter optical-UV continuum and X-rays \citep{Jackson89} and also the near-infrared emission is brighter by a factor of 2-3 \citep{Runnoe2013a}. This anisotropy needs to be understood, and ideally corrected for, in order to obtain the actual bolometric luminosity. Orientation also affects the luminosity functions of quasars \citep{DiPompeo2014}. For these reasons, orientation introduces scatter into the measurement of the fundamental properties of quasars. In order to fully understand the growth of black holes and their relationship to the evolution of galaxies, we require a better understanding and quantification of orientation, an essential parameter for axisymmetric quasars.

\subsection{Orientation indicators for radio-loud quasars}
Quasars are generally too distant to have their inner structure spatially resolved for direct measurements of their orientation
(although see the amazing results for 3C 273 by the \citealt{GRAVITY} and for M87 by the \citealt{EHT}). However, radio-loud quasars have large-extent radio structures that can be used to determine orientation. The unresolved compact bases of these relativistic jets observed as cores undergo Doppler-boosting in the direction of motion \citep{OB82,WBr95,Van2015}. Hence, the core emission is orientation dependent and dominates in face-on quasars. Whereas, the isotropic diffuse emission from lobes dominates in the case of more edge-on quasars. 

The core emission also depends on the power of the central engine. So to account for the power of the central engine, the core emission is normalized, often by the extended emission. Hence, the ratio of the flux density of the core to the extended emission at 5 GHz rest-frame is defined as the core dominance R  \citep{OB82}, and serves as an orientation indicator. But using extended flux density to normalize the central engine power has its own shortcomings. First, extended emission includes the contribution from hotspots, which are formed when highly collimated jets hit material in the intergalactic (or intra-cluster) medium. Hence the extended flux density depends on the gaseous environment in which these quasars reside, which can differ from source to source and with redshift. Second, the extended emission is representative of the time-averaged power of the central engine, i.e., it depends on the history of the source and does not correspond \edit1{directly} to the current power, which governs the core emission.

Likely a better way to normalize the radio core is to use V-band optical luminosity because it is emitted from the contemporaneous accretion of material into the black hole whose rotation powers the jet. Therefore a better quasar orientation indicator is $\rm R_V$, the ratio of radio core luminosity and optical V-band luminosity \citep{WBr95,Van2015}. Evidence supporting the V-band optical luminosity as a better normalization factor includes its correlation with the emission-line luminosity \citep{YO78} and the proportionality of jet power with the luminosity of the narrow-line region \citep{RS91}. Although, the optical emission is anisotropic, between face-on \& edge-on quasars, it varies only by a factor of about two, which is negligible in contrast to a factor of $\sim 10^4$ in the case of core flux density emission. 

Other ways to determine core-dominance are as follows. \cite{RS91} uses luminosity of narrow-line region based on luminosities of [OII] and [OIII] lines to define $\rm R_{NLR}$. \cite{Willott} defines $\rm R_{OX}$ using total luminosity at 151 MHz to normalize the core luminosity. At this frequency, the total luminosity measures the luminosity of lobes. 

\cite{Van2015} performed two tests to determine the best orientation indicator among the four discussed above. The first test uses geometric rank as a proxy for orientation. Each object is assigned a rank as a sum of rank in ascending order of projected linear size and in decreasing order of the angle between the core and two hotspots. They found that only R and $\rm R_v$ shows high correlation with geometric rank. They also performed line regression to fit the data and calculate the data variance. They find that $\rm R_v$ has the lowest variance, i.e., optical luminosity introduces the least amount of scattering in the core-dominance parameter. They conducted another test by randomizing the normalization factor and obtained the distribution of the variance. They shuffled the denominator of each core dominance parameter by the value of the different source and obtained its variance against the best fit line calculated earlier. They repeated this 100,000 times to obtain a distribution of variance. They find that normalizing the core luminosity by a measurement from the same source only matters in the case of $\rm R_v$, when dividing by optical luminosity.  Hence, their work confirms that $\rm R_v$ is a better orientation indicator, as previously claimed by \cite{WBr95}. In this paper, we examine only the extended radio and the optical emission for normalization to determine core dominance. 

%\subsection{A new sample for orientation study}
\subsection{Selection biases in past orientation studies}
Few past orientation studies of radio-loud quasars have employed ideally selected samples. The widely used 3C/3CR catalog at 178 MHz, with a flux limit of 10 Jy, provides a nearly orientation-independent sample, as at this low frequency the emission is dominated by the extended optically thin component. But, the high-flux limit makes this catalog biased towards high-luminosity sources, which may not be typical. \cite{OB82} use a flux-limited sample of 3C quasars, with no redshift constraints, to study their radio core dominance R. Their results on the quasar counts indicate that samples selected at high-frequencies, above 1 GHz, have more high-redshift quasars in comparison to samples selected at lower frequencies of a few hundred MHz. High-redshift quasars will go through higher Doppler-boosting and will have flatter spectra on average. Hence the sample will be biased towards flat-spectrum quasars. 

On the other hand, an optical magnitude limit also introduces a bias towards flat-spectrum radio quasars \citep{JB2013,KS1987}. For e.g., the \cite{WB86} sample is a collection of quasars from different catalogs based on their optical brightness cutoff ($<$17 mag) and a redshift cutoff of 0.7. The anti-correlation between R and optical magnitude \citep{BW1985} means that quasars with jet close to the line of sight have enhanced optical emission. So, their sample includes quasars that are intrinsically fainter than 17 mag but optically beamed. The combination of magnitude, orientation, and redshift results in a sample that lacks faint quasars and is biased against low R quasars. 

\cite{Runnoe2013} studied the effects of orientation on determining the mass of black holes. Their sample consists of 52 radio-loud (RL) quasars from the RL subsample of \cite{Shang2011}, excluding blazars. These objects have extended radio luminosities in a range of $\rm 26.5<log ~L_{ext}<27.6$, measured at 5 GHz in units of $\rm erg~ s^{-1} ~Hz^{-1}$. Hence, the selected objects are similar in nature but differ by their orientation. For our sample, we performed a luminosity cut allowing range in $\rm L_{ext}$ down to 33.0 $\rm erg~ s^{-1} ~Hz^{-1}$ measured at a low frequency (325 MHz) that is dominated by emission from the lobe component.

\subsection{Selection and resolution bias associated with FIRST}
Recent studies are based on samples selected by matching Sloan Digital Sky Survey (SDSS, \citealt{York2000}) and Faint Images of the Radio Sky at Twenty-cm (FIRST, \citealt{Becker}) survey data. A sample selected at high frequency, such as 1.4 GHz in case of FIRST, is expected to be biased toward core-dominated quasars that have flat-spectra in comparison to the less orientation-biased lobe-dominated quasars that have steep-spectra. 

Irrespective of the sample selection, it is quite common to use FIRST measurements to determine core dominance. \cite{JB2013} claim that FIRST measurements, do not have the spatial resolution to reliably distinguish the core from more extended components. So the core flux densities obtained from the FIRST survey will therefore likely have excess flux density coming from the extended emission, thus overestimating the radio core dominance. Thus orientation studies based on FIRST survey radio data are biased by selection and resolution effects (e.g., \citealt{Brotherton2015}).

In order to improve the radio-core dominance measurements, we need high-resolution data that can separate core from extended emission. For instance, for radio data with 5$\arcsec$ resolution, we will not be able to distinguish a 1$\arcsec$ core from any extended emission within the resolution element.  In the absence of high-resolution data, we would then likely over-estimate the core flux densities and poorly identify their orientation. 

To test this idea, we proposed for NSF's Karl G. Jansky Very Large Array (VLA) 10 GHz A-array observations of a well-selected and relatively orientation-unbiased sample of quasars. We selected 10 GHz because at this frequency: measurements are still efficient, and the cores will stand-out against extended emission. Using the A-array can achieve an angular resolution of 0.2", a factor of 25 times smaller than FIRST. From our results, we have confirmed that high-resolution data (better than the $\sim$5$\arcsec$ of FIRST) is generally required to correctly estimate the core flux density of quasars.

The outline of this paper is as follows: Section \ref{sec:sample} discusses the sample selection, followed by Section \ref{sec:obsrvation} about our VLA observations and the reduction procedures used to derive the core flux densities. We catalog all the radio and optical flux densities in Section \ref{sec:data}. Next we define the two core dominance parameters, R and $\rm R_{5100}$ and the groups we determined on the basis of extended radio spectra in Section \ref{sec:R and R5100}. Finally, we present our results in Section \ref{sec:results} followed by discussion and conclusion in Section \ref{sec:discussion_new} and Section \ref{sec:conclusion}. Appendix \ref{sec: app1} shows the spectral energy distribution (SED) of our targets. Appendix \ref{sec: app2} shows radio images of sources with resolved structure at 10 GHz in contrast to their point-like FIRST images. \edit1{It also tabulates the statistical properties of the 10 GHz images presented in this paper. Lastly, Appendix C lists the additional targets we observed with VLA.} For our calculations we have used a cosmology with $\rm H_0 = 70~km~s^{-1}~Mpc^{-1}$, $\rm \Omega_\Lambda = 0.7$ and $\Omega_m = 0.3$. We have defined radio spectral index $\alpha$ by the relation: $\rm S_{\nu}\propto \nu^{\alpha}$, where S stands for the flux density and $\nu$ is frequency.

\section{Sample selection} \label{sec:sample}

We used a relatively unbiased sample (\citealt{JB2020}, ApJ submitted) that is likely largely representative of the full range of quasar orientations. These objects were selected from the Sloan Digital Sky Survey Data Release 7 (SDSS DR7) quasar catalog \citep{Schneider} within a redshift range of $\mbox 0.1<z<0.6$ to attain high-quality optical spectra for the measurements of the $\mbox H\beta$ emission line. The sample was then matched with the low-frequency 325 MHz Westerbork Northern Sky Survey (WENSS, \citealt{WENSS,Rengelink97}), that has a resolution of \edit1{54$\arcsec \times$54$\arcsec$ cosec($\delta$}). WENSS covers the whole sky north of 30$\degree$, about 1/3 of the larger SDSS DR7 quasar catalog covering $\rm \approx 9380 ~deg^2$. A total radio luminosity cutoff $\rm log(L_{325}) > 33.0 ~erg ~s^{-1} ~Hz^{-1}$ was applied. This luminosity cutoff is generally above the WENSS flux limit of 18 mJy ($5\sigma_{rms}$) and also above the transition between Fanaroff-Riley class I and II sources at $\rm log(L_{325}) \sim 32.5 ~erg ~s^{-1} ~Hz^{-1}$.  At 325 MHz, the isotropic extended emission dominates in nearly every object and hence produces a quasar sample largely unbiased by orientation. 

Core and lobe candidates were matched separately because each object may appear as separate entries in the WENSS catalog. Core candidates were matched within a search radius of $\rm 30 \arcsec$, whereas lobe candidates were matched within $1100\arcsec$. To associate lobes with a core candidate three criteria were used; the two lobes should be within 30$\degree$ of being opposite to each other, their ratio of the distance from SDSS position should be within a factor of two and their flux ratio should also be within a factor of two. As a result, 142 sources were selected and visually inspected in SDSS, WENSS, FIRST and the NRAO VLA Sky Survey Catalog (NVSS, \citealt{NVSS}) to ensure correct matching of all the components. Furthermore, 16 sources were removed based on visual inspection of the match at multiple wavelengths and a subsequent re-evaluation of the luminosity cut.

%\deleted{The \citeauthor{JB2020} sample is the basis for the sample used in this work, with some differences.  We report 147 objects in Table 1, 119 objects come from the Runnoe \& Boroson sample and are marked with `RB', and an additional 28 objects are marked with `a' in column 1.  We excluded 7 objects from Runnoe \& Boroson that do not have black hole masses from $M_{\sigma \star}$ (needed for our ultimate science goal of refining orientation indicators).  The additional objects were near the luminosity cut and were proposed for the VLA observations. With the development of Runnoe and Boroson's sample they were latter dropped, but we include them here. We do not generally distinguish these objects in our analysis.}

\begin{figure}[h]
    \centerline{\includegraphics[scale=0.8]{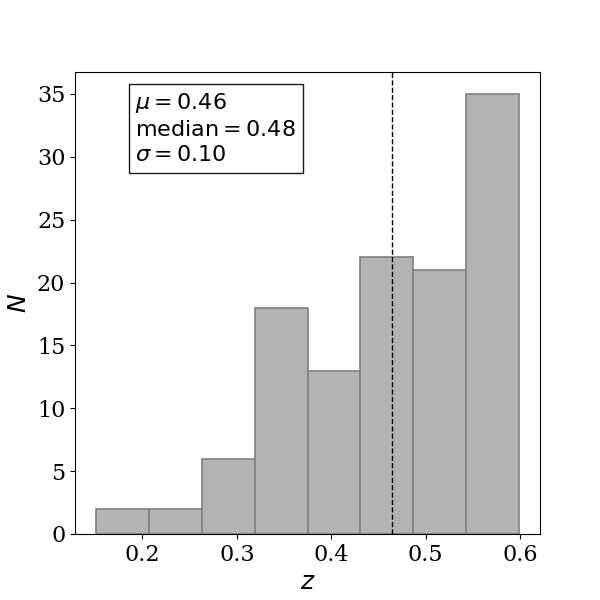}}
    \caption{\edit1{The histogram of redshift distribution of the sample of 119 radio-loud quasars obtained from \cite{JB2020}. The dashed line represents the mean redshift, z = 0.46.\label{fig:fig1}} }
\end{figure}

\edit1{The \citeauthor{JB2020} final sample consists of 126 objects. We excluded seven objects that do not have black hole masses \edit1{from M-$\rm \sigma_{ \star}$ relationship (e.g., \citealt{McConnellandMa2013}), as one of the original goals of our study was to address orientation effect in black hole mass (e.g., \citealt{Brotherton2015}). Here, $\sigma_{\star}$ is the stellar velocity dispersion of the host galaxy and mass estimated using M-$\rm \sigma_{ \star}$ relationship is independent of orientation of quasar axis.}  Those seven objects have problematic optical spectra. Either the broad H$\beta$ line was too weak and a reliable width could not be measured, or the optical continuum was unreliable (e.g., because it was weak). Our final sample consists of 119\footnote{We list additional objects that were proposed for VLA observation but are no longer part of \citeauthor{JB2020} sample in Appendix C.} objects and have a mean redshift of 0.46 (Figure \ref{fig:fig1}). See section 4 of \cite{JB2020} for further discussion on sample properties.}

\section{VLA Observations and Data Reduction} \label{sec:obsrvation}

The VLA continuum observations were carried out at 10 GHz (X-band) with the A-array configuration to achieve 0.2$\arcsec$ resolution, 25 times smaller than FIRST to separate the unresolved core from extended emission. \edit1{The VLA project id of the observations is 16B-329.} We used the 3-bit digital samplers appropriate for X-band wideband observations with basebands of 2x2 GHz and a total bandwidth of 4 GHz. \edit1{The dynamical observations occurred in two scheduling blocks (SB), on  29 October 2016 (SBID 32533323; Total elapsed time $=$ 10508s) and 13 November 2016 (SBID 32713722; Total elapsed time $=$ 9232s).}  For both blocks, the flux calibrator was 3C 286. Each block included a scan of a phase calibrator at the beginning and the end. Some of our targets were already standard phase calibrators, so we also used them for phase calibration. \edit1{The time on source varies between 88s and 286s. The mean time on source is 199s.}
%block1: SBid 32533323  29 OCt 2016 Total elapsed time = 10508 seconds
%block2: SBid 32713722 Total elapsed time = 9232 seconds

Data were reduced using the Common Astronomy Software Applications (CASA, \citealt{CASA}) version 4.7.0. The initial 6 seconds of the data from each scan were flagged to take into account the stabilizing time for the array. We did not flag our data for shadowing, as A-array does not suffer from this effect. Antennas reported with bad baselines in the observer's log were flagged wherever corrections were unavailable. By making plots of amplitude versus time for the flux calibrator we found that the spectral window 12 and 31 in the first block had an error, so we excluded these spectral windows while applying further calibrations. 

The visibilities of the flux calibrator were modeled using the Perley-Butler 2013 \citep{Perley_2013} model for X-band. The initial phase calibration was performed to look for the time variation from scan to scan for the bandpass calibrator before deriving the bandpass solution. Then we solved for the antenna-based delay relative to the reference antenna. These phase and delay solutions were used to determine the bandpass solution that best determines the variation of the gain with frequency. Next, we derived the correction for the complex antenna gain for the flux calibrator and phase calibrators. The absolute flux densities of the phase calibrators were obtained assuming the gain amplitude for the phase calibrators is the same for the flux calibrator, for which we have taken the gain amplitude from the measured and modeled visibilities.  The calibration solutions derived earlier were then applied to the flux and phase calibrators. We matched the targets to the neighboring phase calibrator and applied the solutions. Image synthesis was performed using the {\tt\string CLEAN} task with multi-frequency synthesis imaging mode, Clark (\citeyear{Clark}), as the method of PSF calculation, and the Cotton-Schwab clean algorithm \citep{Schwab} with one term and Briggs weighting \citep{Briggs}.

\section{Essential radio and optical data}\label{sec:data}
The newly synthesized 10 GHz images were visually inspected to ensure matching of the core position, i.e., the position of the pixel with maximum flux density, with the SDSS position. There were \edit1{five} quasars with poor-quality images due to bad radio data \edit1{(J134303.59+521626.7, J154620.98+453916.7, J170441.38+604430.5) or high positional uncertainty (J105232.74+612520.9, J141858.85+394638.7);} that we dropped from further analysis.  We used the CASA task {\tt\string imstat} to measure the 10 GHz core flux densities from the peak flux density at the optical position. The standard deviations in the residual images produced by the {\tt\string CLEAN} task provided the root mean square (RMS) noises.

Table \ref{table1} presents the  properties of our final set of \edit1{119} quasars. Column 1 and 2 lists the SDSS object names and their redshifts, respectively. We list the optical flux densities at 5100 $\mbox{\AA}$ in column 3, obtained from the SDSS spectra of these quasars. We adopted these measurements from \cite{JB2020}. They corrected the spectra for Galactic extinction. \edit1{The model used to fit the AGN power-law continuum also takes into account the contribution from the host galaxy and the optical Fe II.}

Core flux densities at 10 GHz and 1.4 GHz are listed in columns 4 and 5 of Table \ref{table1}, respectively.  For the 1.4 GHz core flux densities we adopt measurements from \cite{JB2020}. They matched to the FIRST survey by manually determined the search radius. They visually determined the core and used the FIRST peak flux. For objects with core detections, they used the peak flux density as the core flux density. For twelve objects with no core detections they used the FIRST RMS and adopted a $\rm 5 \sigma$ limit on the core flux density \edit1{(J080754.50+494627.6, J081540.84+395437.8, J085341.18+405221.7, J090501.55+533907.5, J105141.16+591305.3, J110726.92+361612.2, J134303.59+521626.7, J141054.05+584655.3, J150455.56+564920.3, J151936.72+534255.5, J154620.98+453916.7, J170334.99+391735.6)}. 

The extended, steep-spectrum emission in quasars dominates at MHz frequencies. So, to measure the extended flux densities and the spectral indices of extended emission,\edit1{ we matched our sources with the WENSS (discussed in Section \ref{sec:sample}), 7th Cambridge (7C, \citealt{7C}), TIFR GMRT Sky Survey Alternative Data release (TGSS ADR, \citealt{Intema2017}) and VLA Low-Frequency Sky Survey redux (\footnote{https://www.cv.nrao.edu/vlss/VLSSlist.shtml}VLSSr, \citealt{Lane2012, Lane2014}) surveys. The total flux densities are listed in column 6, 7, 8 and 9 of Table \ref{table1}, respectively. We report the relative error in flux densities of WENSS using equation 9 of the catalog paper. The 7C survey covers an area of about 1.7 sr at 151 MHz and has a resolution of 70$\arcsec \times$ 70$\arcsec$ cosec$\rm (\delta)$. We used the formula given in the appendix of the catalog paper to convert the cataloged signal to noise ratio to the RMS (1$\sigma$) error in the flux densities. TGSS ADR is a recently available radio survey at 150 MHz with 25$\arcsec$ resolution and covers 90\% of the full sky with median RMS noise of 3.5 mJy beam$^{-1}$. The catalog also reports the flux density error for each target. For both the surveys, we started with a search radius of 70$\arcsec$. We then visually matched images with WENSS and subsequently increased the search radius to 5$\arcmin$ for some targets to collect all the components. We used the same approach to match our targets with VLSSr survey. It is a low frequency counterpart of the NVSS at 74 MHz with 75$\arcsec$ resolution and $\approx$0.1 Jy beam$^{-1}$ RMS sensitivity. We report the cataloged error estimates in flux densities.} The radio spectral index of extended emission, $\alpha_{ext}$, is listed in column 10. \edit1{Last column lists the group assigned to these objects on the basis of their radio SED. We will address the reason for choosing two surveys at 150 MHz, the method used to calculate $\alpha_{ext}$ and the group assignment in the next section.}

\begin{longrotatetable}
\begin{deluxetable*}{ccccccccccc}
\tablecaption{SDSS names, redshifts, optical and radio flux densities, and spectral indices of extended emission of 119 radio-loud quasars.\label{table1}}
\tabletypesize{\scriptsize}
\tablehead{
\colhead{Object name} & \colhead{Redshift} & \colhead{\tablenotemark{$_1$}F5100} & \colhead{$\rm S^{core}_{10 GHz}$} & \colhead{$\rm S^{core}_{1.4 GHz}$} & \colhead{$\rm S_{WENSS}^{total}$} & \colhead{$\rm S_{TGSS}^{total}$} & \colhead{$\rm S_{7C}^{total}$} & \colhead{$\rm S_{VLSSr}^{total}$} & \colhead{$\rm \alpha_{ext}$}  & \colhead{\tablenotemark{$_2$}Group}\\
\colhead{SDSS} & \colhead{z} & \colhead{} & \colhead{$(\rm mJy~beam^{-1})$} & \colhead{$(\rm mJy~beam^{-1})$} & \colhead{$(\rm mJy$)} & \colhead{$(\rm mJy$)} & \colhead{$(\rm mJy$)}  & \colhead{$(\rm mJy$)} & \colhead{} & \colhead{}
}
\colnumbers
\startdata
% J073422.19+472918.8 	&	0.382	&	9.67$\pm$0.25   	&	12.39$\pm$0.06   	&	6.47$\pm$0.14    	&	766$\pm$31     	&	991.9$\pm$99.8     	&	1257$\pm$87    	&	2020$\pm$300   	&	-0.71$\pm$0.049  	\\
% J074125.22+333319.9 	&	0.364	&	19.38$\pm$0.22  	&	8.61$\pm$0.03    	&	3.01$\pm$0.22    	&	1853$\pm$74    	&	3683.1$\pm$261.3   	&	2789$\pm$146   	&	5330$\pm$470   	&	-0.754$\pm$0.04  	\\
% J074541.66+314256.6 	&	0.461	&	96.02$\pm$0.12  	&	449.8$\pm$0.42   	&	614.64$\pm$0.15  	&	3458$\pm$138   	&	6802$\pm$484.7     	&	5948$\pm$296   	&	15370$\pm$1340 	&	-0.716$\pm$0.045 	\\
% J075145.14+411535.8 	&	0.429	&	5.24$\pm$0.29   	&	94.16$\pm$0.41   	&	202.68$\pm$0.14  	&	460$\pm$19     	&	533.3$\pm$53.6     	&	457$\pm$48     	&	730$\pm$100    	&	-0.748$\pm$0.105 	\\
% J080413.87+470442.8 	&	0.510	&	2.30$\pm$0.29   	&	107.5$\pm$0.43   	&	847.18$\pm$0.21  	&	2574$\pm$103   	&	4586.6$\pm$459.5   	&	3644$\pm$184   	&	3720$\pm$450   	&	-0.64$\pm$0.049  	\\
% J080644.42+484149.2 	&	0.370	&	23.09$\pm$0.27  	&	81.08$\pm$0.25   	&	43.32$\pm$0.14   	&	3099$\pm$124   	&	6867.8$\pm$519     	&	6071$\pm$298   	&	1210$\pm$160   	&	-0.766$\pm$0.044 	\\
% J080754.50+494627.6 	&	0.575	&	2.65$\pm$0.29   	&	1.7$\pm$0.02     	&	1.05$\pm$0.21    	&	1452$\pm$59    	&	3061.4$\pm$306.5   	&	2787$\pm$157   	&	5000$\pm$630   	&	-0.897$\pm$0.051 	\\
% J080814.70+475244.7 	&	0.546	&	14.00$\pm$0.26  	&	3.26$\pm$0.03    	&	26.39$\pm$0.15   	&	226$\pm$11     	&	365.4$\pm$41.9     	&	610$\pm$76     	&	540$\pm$100    	&	-0.827$\pm$0.08  	\\
J073422.19+472918.8 & 0.382 & 9.67$\pm$0.25 & 12.39$\pm$0.06 & 6.47$\pm$0.14 & 766$\pm$31 & 991.9$\pm$99.8 & 1257$\pm$87 & 2020$\pm$300 & -0.71$\pm$0.049 & 2 \\
J074125.22+333319.9 & 0.364 & 19.38$\pm$0.22 & 8.61$\pm$0.03 & 3.01$\pm$0.22 & 1853$\pm$74 & 3683.1$\pm$261.3 & 2789$\pm$146 & 5330$\pm$470 & -0.754$\pm$0.04 & 2 \\
J074541.66+314256.6 & 0.461 & 96.02$\pm$0.12 & 449.8$\pm$0.42 & 614.64$\pm$0.15 & 3458$\pm$138 & 6802$\pm$484.7 & 5948$\pm$296 & 15370$\pm$1340 & -0.716$\pm$0.045 & 2 \\
J075145.14+411535.8 & 0.429 & 5.24$\pm$0.29 & 94.16$\pm$0.41 & 202.68$\pm$0.14 & 460$\pm$19 & 533.3$\pm$53.6 & 457$\pm$48 & 730$\pm$100 & -0.748$\pm$0.105 & 3 \\
J080413.87+470442.8 & 0.510 & 2.30$\pm$0.29 & 107.5$\pm$0.43 & 847.18$\pm$0.21 & 2574$\pm$103 & 4586.6$\pm$459.5 & 3644$\pm$184 & 3720$\pm$450 & -0.64$\pm$0.049 & 2f\\
J080644.42+484149.2 & 0.370 & 23.09$\pm$0.27 & 81.08$\pm$0.25 & 43.32$\pm$0.14 & 3099$\pm$124 & 6867.8$\pm$519 & 6071$\pm$298 & 1210$\pm$160 & -0.766$\pm$0.044 & 2 \\
J080754.50+494627.6 & 0.575 & 2.65$\pm$0.29 & 1.7$\pm$0.02 & 1.05$\pm$0.21 & 1452$\pm$59 & 3061.4$\pm$306.5 & 2787$\pm$157 & 5000$\pm$630 & -0.897$\pm$0.051 & 2 \\
J080814.70+475244.7 & 0.546 & 14.00$\pm$0.26 & 3.26$\pm$0.03 & 26.39$\pm$0.15 & 226$\pm$11 & 365.4$\pm$41.9 & 610$\pm$76 & 540$\pm$100 & -0.827$\pm$0.08 & 2 \\
\enddata
\tablenotetext{1}{F5100 is in units of $\rm 10^{-17} erg~ s^{-1} \AA^{-1} cm^{-2}$.}
\tablenotetext{2}{Quasars are assigned group 1, 2, 2-flag (2f) or 3 on the basis of their radio spectrum. See section 5 for details.}
\tablecomments{ Table 1 is published in its entirety in the machine readable format.  A portion is shown here for guidance regarding its form and content.}
\end{deluxetable*}
\end{longrotatetable}

\section{Core dominance determinations} \label{sec:R and R5100}
We derived two radio orientation indicators: (1) R, radio core dominance, defined as the ratio of the observed core flux density and the observed extended flux density k-corrected to 5 GHz rest-frame (equation 1), (2) $\rm R_{5100}$ defined as the ratio of the core flux density and the optical 5100 $\mbox \AA$ flux density (equation 2).We derive both orientation indicators using core flux densities at observed frequencies, $\rm \nu_{obs}$, of 1.4 GHz (from FIRST) and 10 GHz.

\begin{equation}
\rm R_{\nu_{obs}} =  (S_{\nu_{obs}}^{core} / S_{obs}^{ext}) |_{5GHz, rest}
\end{equation}

\begin{equation}
\rm R_{5100,\nu_{obs}} =  S_{\nu_{obs}}^{core} / S_{5100\text{\AA}}
\end{equation}

The flux densities ($\rm S_{obs}$) observed at frequency $\rm \nu_{obs}$ are k-corrected to 5 GHz rest-frame using
\begin{equation}
\rm S_{5 \rm GHz,rest} =  S_{obs}~ (1+z)^{-(1+\alpha)} \left( \frac{5 \rm ~GHz}{\nu_{\rm obs}} \right)^{\alpha},
\end{equation}
where, z is the redshift and $\alpha$ is the radio spectral index. We assumed the radio spectral index of the cores to be zero, as quasar cores have flat-spectra (\citealt{Bridle}, \citealt{Kimball2011}), so,  $\rm S^{core}_{5 GHz, rest} = S^{core}_{\nu_{obs}} / (1+z)$. The observed extended flux density is the difference of total flux density at the observed frequency ($\rm \nu^\prime_{obs}$) and the core flux density at the observed frequency ($\rm \nu_{obs}$) i.e. $\rm S^{ext}_{obs} = S^{total}_{\nu^\prime_{obs}} - S^{core}_{\nu_{obs}}$. To calculate $ \rm S^{ext}_{5GHz,rest}$, we need to measure the extended radio spectral index, $\alpha_{ext}$.

The best way to calculate $\rm \alpha_{ext}$ is to fit a power-law slope to the total flux densities obtained from simultaneous observations at two low radio frequencies, where the emission is dominated by the lobes. In practice, observations are not available at low enough frequency and one still needs to know whether an object is core-dominated or lobe-dominated. If an object is core-dominated, total flux density will be dominated by the core emission rather than the lobe emission, giving an inaccurate measure of $\rm \alpha_{ext}$. So, to determine the most accurate $\alpha_{ext}$ we plotted the radio spectral energy distribution (SED) of our targets. We plotted the total flux densities from \edit1{VLSSr}, TGSS, 7C, WENSS, NVSS, and FIRST. \edit1{The NVSS and FIRST total flux densities are taken from Table 1 of \cite{JB2020}}. We also included the core flux densities from FIRST and the new 10 GHz observations (Appendix \ref{sec: app1}). By visually inspecting these radio SEDs we were able to distinguish flat-spectrum sources (core-dominated) from the steep-spectrum sources (lobe-dominated). 
\begin{figure}[h]
    \centerline{\includegraphics[scale=0.5]{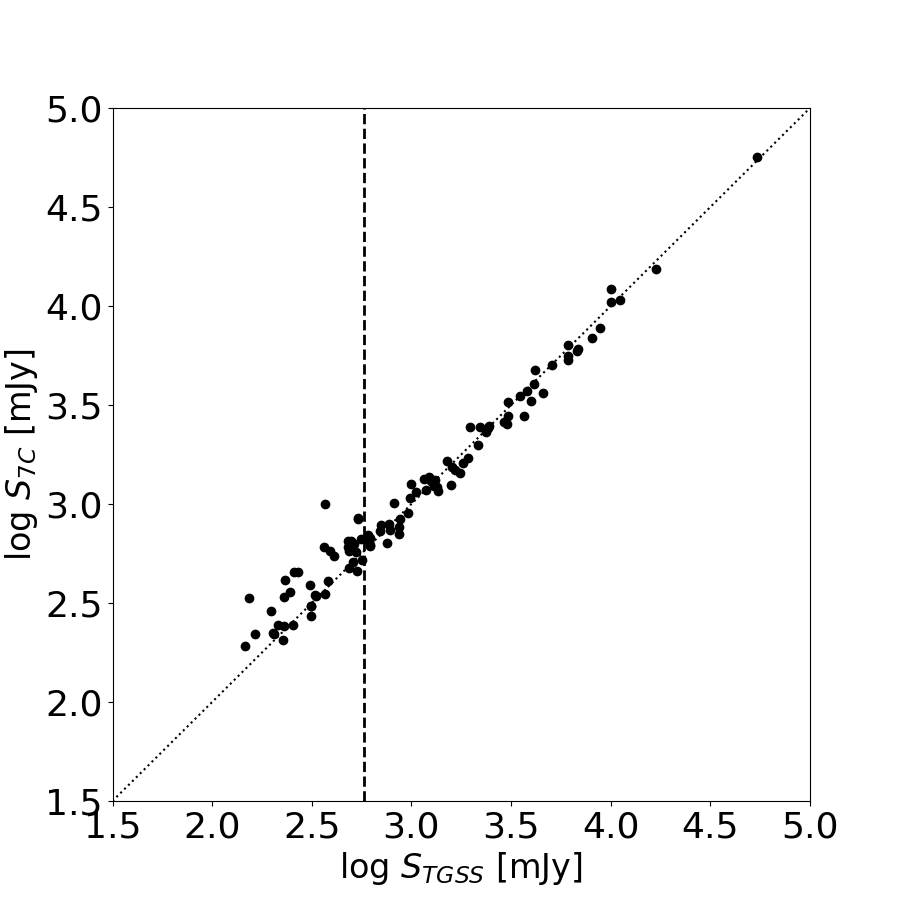}}
    \caption{The total flux densities from 7C (y-axis) versus TGSS (x-axis) in log scale. The dotted line represents the 1:1 ratio and the dashed line represents the cut at \edit1{$\rm S_{TGSS} = 580$ mJy} below which we disregard TGSS values as they suffer from missing extended components.\label{fig:fig2}}
\end{figure}

For a significant number of quasars, we found that the total flux densities from TGSS were falling below the overall slope (eg., J080814.70+475244.7, J080833.36+424836.3, J085341.18+405221.7, J110313.30+301442.7, etc.). This is the reason why we included the 7C survey in our analysis, despite being at a frequency so close to that of TGSS. Figure \ref{fig:fig2} compares the total flux densities from TGSS and 7C in log scales. The plot shows an agreement between the two surveys only for  \edit1{$\rm S_{TGSS} \geq 580$ mJy}. The difference between TGSS and 7C at lower flux densities is because of the resolution bias, 7C having a larger beam of 70$\arcsec$. The ratio of flux densities of the two surveys matches to almost unity for sources brighter than 2 Jy (see section 4.5 of \citealt{Intema2017}). We therefore made a cut at \edit1{$\rm S_{TGSS} = 580$ mJy} and discarded TGSS measurements that are below the cut while plotting radio SEDs. The total flux densities between WENSS, 7C, TGSS and VLSSr agree well with a consistent power-law in a majority of our sources. 

\begin{figure}
    \centerline{\includegraphics[scale=0.7]{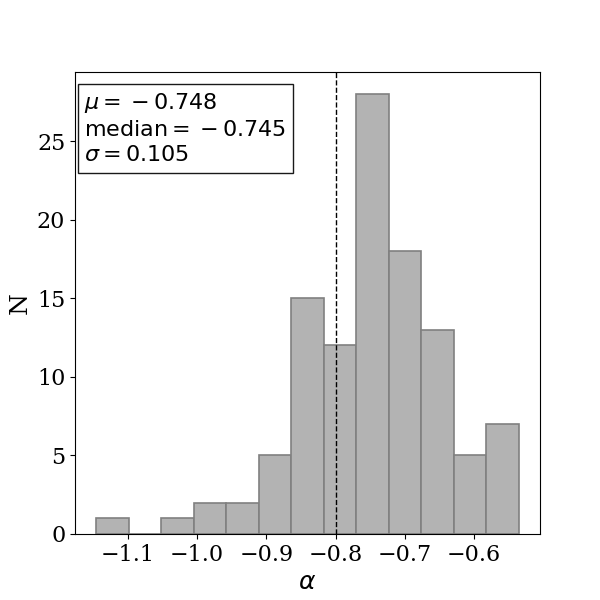}}
    \caption{Histogram of $\rm \alpha_{ext}$ measured from the slope of the best-fit line to the radio spectra of quasars in group 2 and group2-flag, that are possibly lobe-dominated quasars. The dashed line at $-0.8$ represents the typical value for $\rm \alpha_{ext}$ (\citealt{Kimball2011}, \citealt{KI2008}). The mean of our sample lies nearby at $\rm \alpha_{ext} = -0.748$. \label{fig:fig3}}
\end{figure}

We measured $\rm \alpha_{ext}$ by fitting a slope to the radio spectra between 150 MHz and 1.4 GHz including total flux densities from \edit1{VLSSr}, TGSS, 7C, WENSS, and NVSS. Although being at 1.4 GHz, NVSS flux density for core-dominated objects is contaminated by core emission (eg., J084957.97+510829.0, J123413.33+475351.2, etc.) and not ideal for $\rm \alpha_{ext}$ measurement, it can be used for lobe-dominated objects as it follows the overall extended slope.  We included NVSS so that we can measure $\rm \alpha_{ext}$ for targets that do not have 7C, TGSS and \edit1{VLSSr} observations. NVSS also served as an reference point along with WENSS \edit1{and VLSSr} in the SEDs to identify the resolution bias in TGSS for fainter targets. We used the measured $\rm \alpha_{ext}$ to differentiate between core-dominated, lobe-dominated, and intermediate quasars. Only for the lobe-dominated quasars, we use the measured value of $\rm \alpha_{ext}$ for our calculations. For core-dominated and intermediate quasars we use the sample mean of $\rm \alpha_{ext}$.

On the basis of $\rm \alpha_{ext}$ and its error we divided our sample into three groups and one sub-group. Group1 consists of \edit1{three} sources with flat radio spectra over 150 MHz to 10 GHz, indicating they are likely core-dominated quasars. Group2 consists of  \edit1{109} sources with $\rm \alpha_{ext}$ steeper than $-$0.5 and very small errors. These sources are likely lobe-dominated quasars.   \edit1{We made a separate sub-group, Group2-flag, of 18 sources with $\rm \alpha_{ext}$ steeper than $-0.5$ and either the same FIRST core flux densities and NVSS total flux densities or comparable FIRST core flux densities and NVSS total flux densities with $\rm \alpha_{ext}$ error greater than 10\%. We suspect some of them to be compact steep-spectrum (CSS) sources.} Figure \ref{fig:fig3} shows the distribution of $\rm \alpha_{ext}$ in the quasars of Group2 and Group2-flag and shows a mean $\rm \alpha_{ext}$ of \edit1{$-0.748$}. Finally, Group3 consists of \edit1{seven} sources with $\rm \alpha_{ext}$ flatter than $-$0.5 and errors greater than 10\%. These sources are likely intermediate between core and lobe-dominated quasars. 

For Group2 and Group2-flag, we calculated $\rm S_{obs}^{ext}$ taking the difference between the total flux densities at $\nu^{\prime}_{obs} = 325$ MHz (from WENSS) and core flux densities at $\nu _{obs}=$ 1.4 GHz (10 GHz). We used the measured $\rm \alpha_{ext}$ to k-correct the extended flux density to the 5 GHz rest-frame (equation 3) and calculated $\rm R_{1.4GHz}~ (R_{10GHz})$ using equation 1.  For Group1 and Group3, consisting of quasars that are likely core dominated, we ideally needed simultaneous measurement of core and total flux density to calculate the extended flux density. Therefore, we used the difference between the FIRST total and core flux densities to measure $\rm S_{obs}^{ext}$. We used the mean $\alpha_{ext}$ of the rest of our sample, i.e., $-$0.748 (Figure \ref{fig:fig3}), to k-correct the extended flux densities to 5 GHz rest-frame. Table \ref{tab:table2} lists the logarithm of radio core dominance evaluated at $\nu_{obs}$ of 1.4 GHz (log $\rm R_{1.4 GHz}$) and 10 GHz (log $\rm R_{10 GHz}$) in column 3 and column 4, respectively. We used the 5100 $\mbox \AA$ flux densities to normalize the radio core flux densities at the same $\nu_{obs}$. The log $\rm R_{5100, 1.4 GHz}$ and log $\rm R_{5100, 10 GHz}$ are given in columns 5 and 6.

\startlongtable
\begin{deluxetable*}{crrcc}
\tablecaption{ Logarithm of $\rm R$ and $\rm R_{5100}$ calculated using new 10 GHz and FIRST 1.4 GHz core flux densities. \label{tab:table2}}
\tablehead{
\colhead{Object name} & \colhead{log($\rm R_{1.4 GHz}$)} & \colhead{log($\rm R_{10 GHz}$)} & \colhead{log($\rm R_{5100, 1.4 GHz}$)} & \colhead{log($\rm R_{5100, 10 GHz}$)}
}
\colnumbers
\startdata
J073422.19+472918.8	&	-1.33$\pm$0.056	&	-1.04$\pm$0.052	&	1.89$\pm$0.034	&	2.17$\pm$0.026\\
J074125.22+333319.9	&	-2.00$\pm$0.084	&	-1.54$\pm$0.042	&	1.25$\pm$0.074	&	1.71$\pm$0.012\\
J074541.66+314256.6	&	0.07$\pm$0.046	&	-0.09$\pm$0.046	&	2.87$\pm$0.001	&	2.73$\pm$0.002\\
J075145.14+411535.8	&	1.26$\pm$0.043	&	0.93$\pm$0.043	&	3.65$\pm$0.055	&	3.32$\pm$0.055\\
J080413.87+470442.8	&	0.34$\pm$0.049	&	-0.72$\pm$0.050	&	4.63$\pm$0.124	&	3.73$\pm$0.124\\
J080644.42+484149.2	&	-1.04$\pm$0.046	&	-0.77$\pm$0.046	&	2.33$\pm$0.012	&	2.61$\pm$0.012\\
J080754.50+494627.6	&	-2.25$\pm$0.206	&	-2.04$\pm$0.052	&	1.66$\pm$0.228	&	1.87$\pm$0.109\\
\enddata
\tablecomments{ Table 2 is published in its entirety online in machine readable format. A portion is shown here for guidance regarding its form and content.}
\end{deluxetable*}

\begin{figure*}[p]
\epsscale{0.83}
\plotone{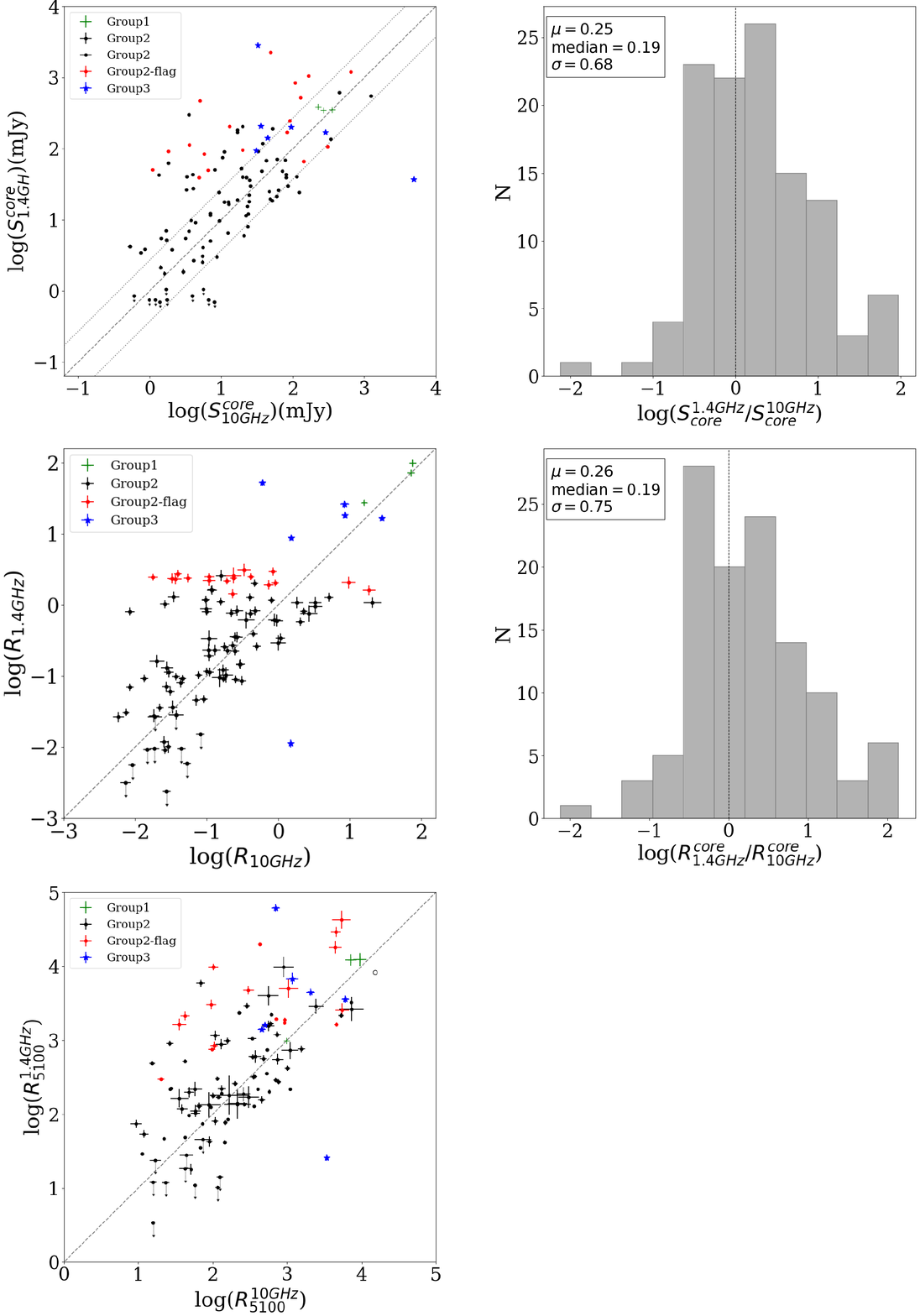}
\caption{The left panels show scatter plots of core flux density measured at 1.4 GHz and 10 GHz and the corresponding R and $\rm R_{5100}$ measurements on log scales. The points are color coded according to the groups defined by the radio spectra of their extended emission. Points on the bottom left corner panel with downward arrows represent the objects with FIRST flux limits. The dashed lines show where the two quantities are equal. \textbf{Top left:} A scatter plot with the logarithm of 1.4 GHz core flux density on y-axis and the logarithm of 10 GHz core flux density on the x-axis.  The dotted line represents $\rm y= x \pm 0.43$, an estimate of the $1\sigma$ scatter introduced by the core variability. \textbf{Middle left:} scatter plot of log R calculated from the 1.4 GHz FIRST cores on the y-axis and from the 10 GHz cores on the x-axis normalized by the extended flux densities scaled to 5 GHz rest-frame according to the groups as discussed in section \ref{sec:R and R5100}. \textbf{Bottom left:} scatter plot of $\rm R_{5100}$ values calculated from the ratio of 10 GHz vs 1.4 GHz core flux densities and 5100$\mbox{\AA}$ flux density. The point with black open circle (\edit1{top right}) represent a quasar with a bad 5100 \AA\ flux density measurement.
{\bf Right:} panels show the histograms of the corresponding ratio of the quantities in the x and y axis of the left panels. The dashed lines in these histograms show where the ratios are equal. \label{fig:fig4}}
\end{figure*}

\section{Results} \label{sec:results}

Figure \ref{fig:fig4} compares the core flux densities at 1.4 GHz (from FIRST) and 10 GHz (top panel) and their impact on the determination of the radio core dominance, R (middle panel) and $\rm R_{5100}$ (bottom panel). The radio cores of quasars typically have a flat spectrum over a broad range of frequencies, so we expect that the core flux density at the FIRST and the observed frequencies to be the same. There are three factors that can introduce differences: variability, inverted core spectrum, and insufficient resolution.

The scatter plot of 1.4 GHz vs 10 GHz core flux densities provides a path to statistically estimate the core variability for our sample. The objects below the 1:1 line have 10 GHz core flux density larger than the 1.4 GHz flux density (except the points with arrows that represent the FIRST detection limit). This increase in the core flux density is likely due to core variability. In order to quantify the variability level, we plot the histogram of the logarithm of the ratio of the two flux densities on the top right panel of Figure \ref{fig:fig4}. The distribution is asymmetric around the dashed line representing equal fluxes. \edit1{A Gaussian fit to the distribution has a mean of 0.25 and a standard deviation of 0.68.}. The core variability contributes to the negative tail of the distribution. \edit1{The standard deviation of a symmetric Gaussian around the dashed line fitting the negative tail is 0.43. So, a change of factor of $\sim$2.69 in the core flux density can be attributed to the core variability}. \edit1{This is an upper limit as the objects with FIRST core flux density limit are also included. Without them, the standard deviation is 0.35 implying a factor of $\sim$2.24 core variability.}

The assumption of a flat radio core spectra could introduce scatter around the 1:1 line that represents $\rm \alpha_{core} = 0$. \cite{MOJAVE} found the mean $\rm \alpha_{core} = 0.22\pm 0.04$ for a sample of 133 flat-spectrum radio quasars using VLBA observations. Their mean core spectral index also shows an anti-correlation with the linear size of the core, implying that the measurements for high-redshift sources were contaminated by the jet emission. To quantify the effect of inverted core spectral index we generated 10,000 samples of  $\rm \alpha_{core}$ from a Gaussian distribution with a mean of 0.22 and standard deviation of 0.04 and calculated the mean flux density of the sample scaled between 1.4 GHz and 10 GHz. We found that the change in flux density is a factor of 1.54 due to non-zero core spectral index. This is the largest effect that could be induced due to an inverted core spectral index which is still smaller than the apparent factor of \edit1{$\sim$2.69} resulting from core variability. 

The scatter plot of 1.4 GHz vs 10 GHz core flux densities also shows that majority of our targets are above the 1:1 dashed line, indicating that many \edit1{FIRST} core measurements are contaminated by poorly resolved extended emission. We found \edit1{41} objects for which FIRST fluxes are above the dotted line representing the core variability limit\edit1{, this is 36\% of the sample. Excluding the targets within the FIRST core flux density limit, the percentage increases to 41\%.} In a few of them we have detected extended features in the 10 GHz images at the scale of a few arcseconds or less (see Appendix \ref{sec: app2}).

%Above edit still correct meaning?  MSB

\begin{figure}[h!]
    \centerline{\includegraphics[scale=0.4
    ]{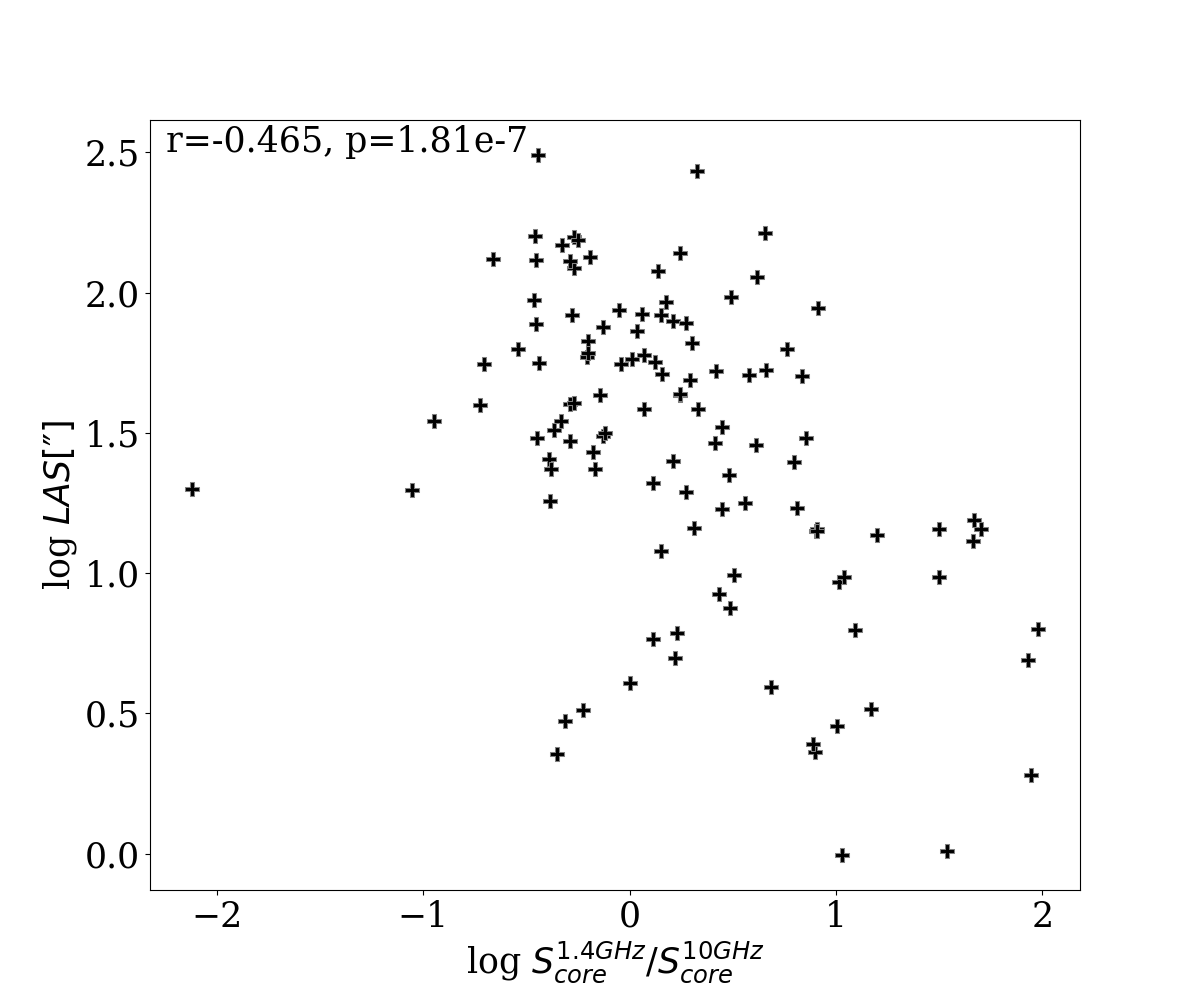}}
    \caption{An anti-correlation between the ratio of the flux densities at 1.4 GHz (from FIRST) and 10 GHz (x-axis) vs the largest angular size (LAS) (in units of arcseconds) from FIRST in log-log scale (y-axis).  Pearson's correlation coefficient is r = $-$0.465 corresponding to a probability that these quantities are uncorrelated of 1.81e-7.\label{fig:fig5}}
\end{figure}

If the FIRST core measurements are contaminated by the extended emission, we expect larger contamination as the angular size of the target in FIRST image decreases. Figure \ref{fig:fig5} plots the largest angular size (LAS) from FIRST against the ratio of flux densities at 1.4 GHz (from FIRST) and 10 GHz in log-log scale. The largest angular size were taken from Table 1 of \cite{JB2020}. It was determined by plotting the FIRST elliptical Gaussian fits to all components associated with a source and calculating the largest associated size scale among them. We find that the ratio increases as the FIRST angular size decreases. The anti-correlation has a Pearson's r-coefficient of -0.465.

These results clearly demonstrate that the cores were unresolved at the FIRST resolution and their core flux density measurements were contaminated with extended emission. With our new X-band observations, we were able to resolve these sources and hence improved the core flux density measurements.

Since the core flux densities were systematically higher in the case of the FIRST measurements, we expect the radio core dominance calculated from the FIRST survey to have higher values too. Figure \ref{fig:fig4} (middle panel) compares R using the FIRST core and 10 GHz core measurements and shows a prominent excess. We observed the same excess in the $\rm R_{5100}$ in figure \ref{fig:fig4} (bottom panel) while using FIRST core measurements. We have plotted the error bars accounting for uncertainties in flux densities and $\alpha_{ext}$. \edit1{The histogram of ratio of $\rm R_{5100}$ from FIRST and 10 GHz is same as ratio of core flux densities from them.}

Our results confirm the \cite{JB2013} claim about resolution effects, and that radio core dominance calculated from FIRST maps is often compromised and biased high. With our current high-resolution radio images, we have determined improved core flux densities and hence improved the radio orientation indicators R and $\rm R_{5100}$. 

\section{Discussion} \label{sec:discussion_new}

There are a number of choices to make when determining the radio core dominance of a quasar.
Observations are conducted at a particular frequency and resolution given available facilities and/or surveys.  There are distinctly different issues when it comes to measuring radio core flux density compared to that of the total or extended radio emission.

The radio core stands out against extended emission best at higher frequencies, making 10 GHz a better choice compared to lower frequencies.  Similarly, radio cores are better distinguished from extended emission at higher angular resolution.  These considerations make our choice of VLA A-array 10 GHz observations among the best to measure quasar radio cores.  As previously discussed, the FIRST Survey using the VLA at B-array and 1.4 GHz suffers shortcomings that can compromise the results for individual objects.

\begin{figure}[h!]
    \gridline{\fig{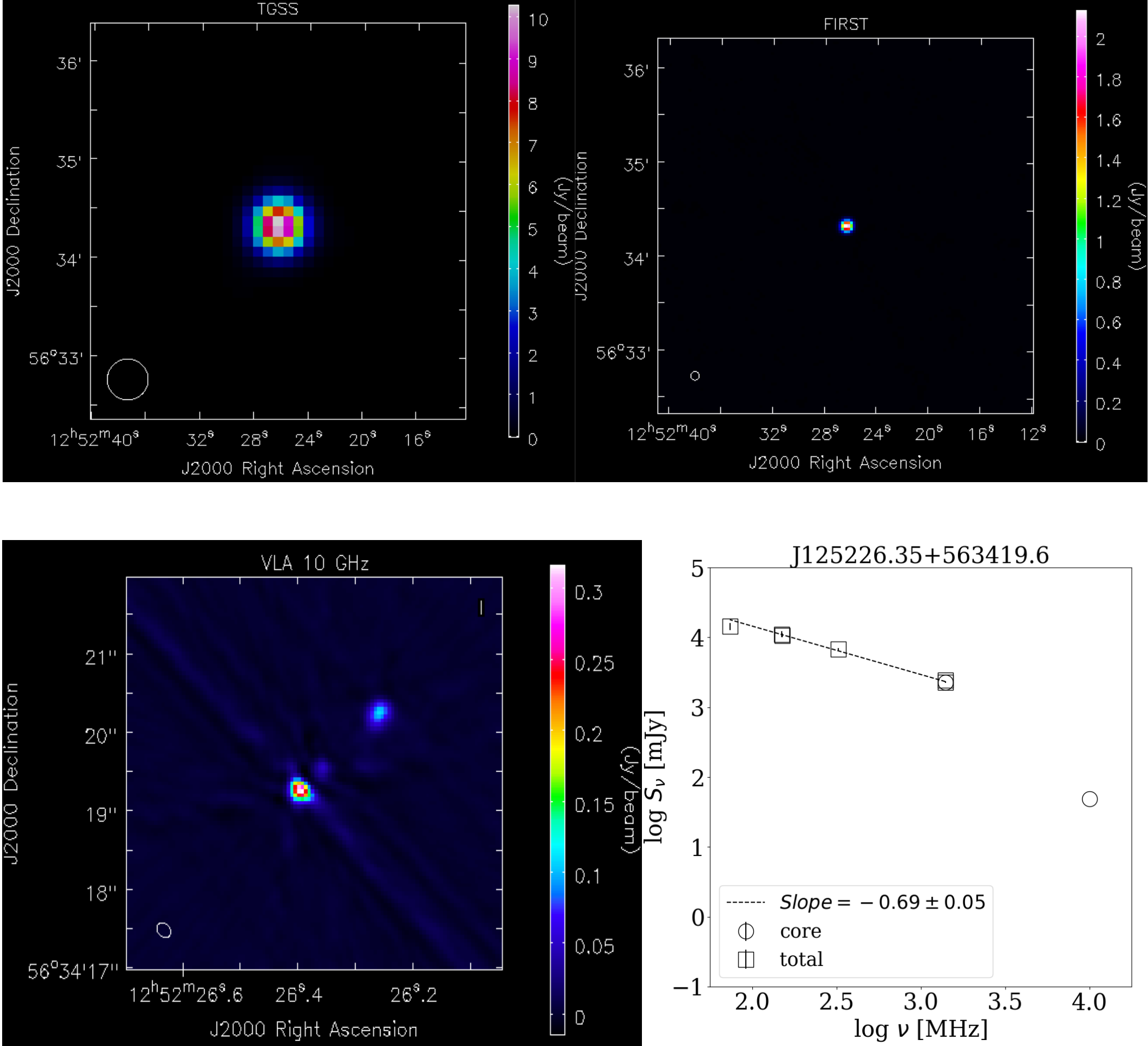}{0.6\textwidth}{(a) J125226.35+563419.6}}
    %\centerline{\includegraphics[scale=0.6]{J125226-grid.pdf}}
    \gridline{\fig{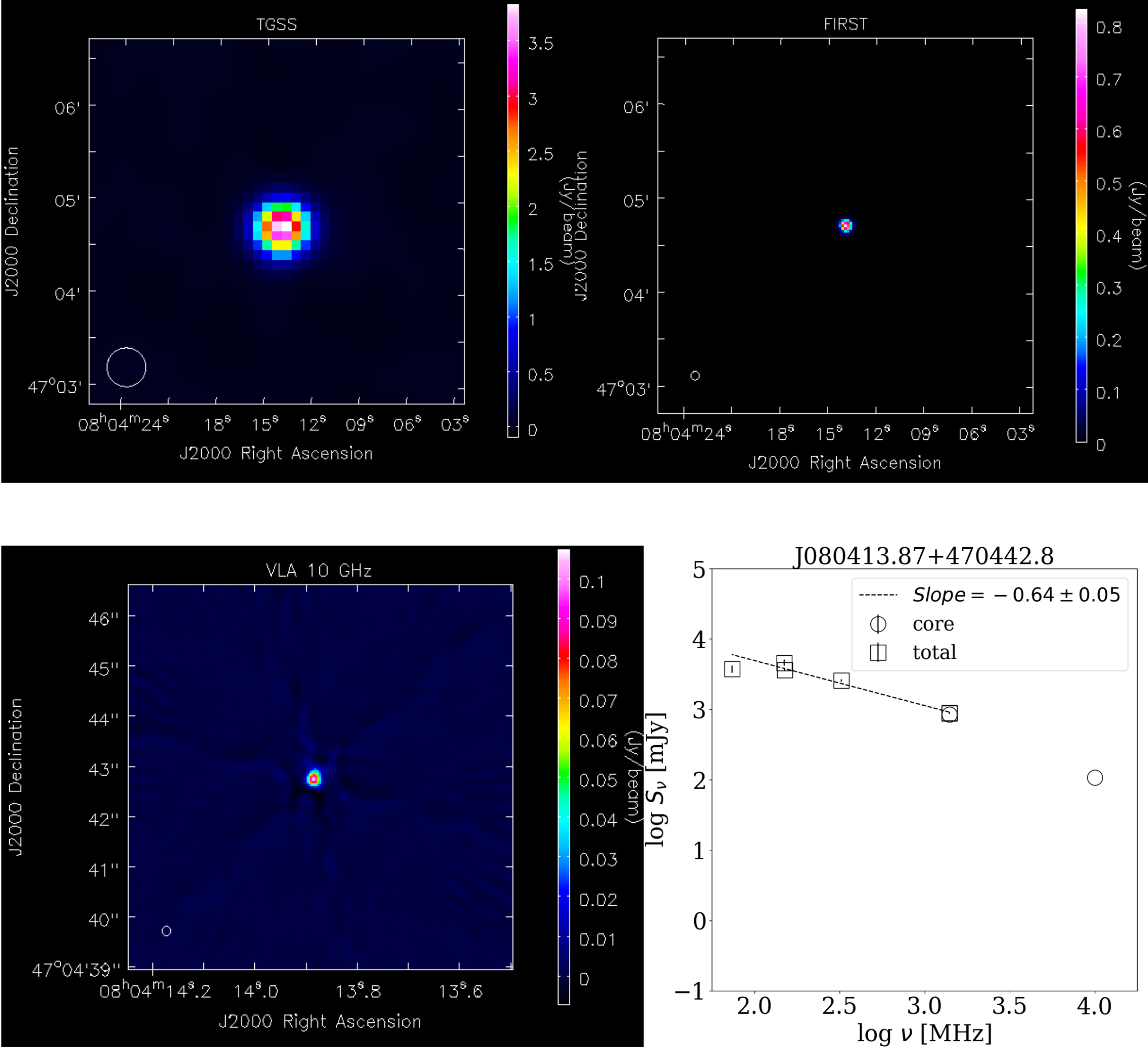}{0.6\textwidth}{(b) J080413.87+470442.8}}
    \caption{TGSS, FIRST and VLA 10 GHz images (see Table 3 for image statistics) of two known CSS sources in our sample that are either (a) resolved or (b) unresolved at 10 GHz. In both cases the contamination of core flux density due to extended emission is removed in the 10 GHz observation. Their SEDs show VLSSr flux density (at 74 MHz) falls below the steep slope, expected in case of CSS quasars. \label{fig:fig6}}
\end{figure}

\edit1{The core flux density of the Group2-flag objects suffers most due to the resolution effect. These sources appear as point-like sources in both GHz and MHz frequency surveys. One can misclassify them as core-dominated sources, but they are compact steep-spectrum (CSS) quasars. The CSS sources are likely young AGNs that have an intrinsically compact radio structure and exhibit a steep radio spectrum that peaks around 100 MHz \citep{Odea1998}. For example, J080413.87+470442.8, J081253.10+401859.9, J115727.60+431806.3, J125226.35+563419.6 and J164311.34+315618.4 are classified as CCS quasars in previous studies such as \cite{Fanti2001} and \citealt{Mai2020}. Figure 6a shows CSS quasar J125226.35+563419.6 appearing as a point source in the TGSS, the highest resolution MHz frequency survey, and in the FIRST images. Using the new 10 GHz VLA observations, we were able to resolve this source. Its SED demonstrates a steep radio slope that likely peaks around 100 MHz. There are other CSS sources like J080413.87+470442.8 that are point source in TGSS and FIRST, and have steep-spectra but the 10 GHz VLA observation was not able to resolve inner structure. At 10 GHz, the core stands out as the contribution from the extended emission decreases. Its SED shows the flattening of radio spectrum above 100 MHz. The CSS quasars are also known to show core-jet structures \cite{Odea1998}. Four Group2-flag objects (J083956.55+422755.8, J095820.94+322402.2, J110313.30+301442.7 and J165005.48+414032.4) are classified as compact symmetric objects (CSO) or have jets in their VLBA images \citep{Helmboldt2007}. A similar jet-like structure is seen in the 10 GHz VLA image of J085128.92+600320.1 (Appendix \ref{sec: app2}), another target in Group2-flag. }

In the future, surveys like the VLA Sky Survey (VLASS) will provide a better source of archival core flux densities than the FIRST. VLASS will cover the entire sky above $-$40 degrees at 2-4 GHz with a resolution of 2.5 arcsec (\citealt{VLASS}). This will provide the radio cores at double the frequency and resolution of the FIRST. There may still be issues of core contamination, but they will be much reduced, although A-array 10 GHz observations remain the optimal choice when possible.

Even after making the best choices for measuring radio cores, our measurements are affected by intrinsic core variability. The cores exhibit a variability factor of up to about two over timescales of a few years (\citealt{Barthel}, \citealt{Kellermann}). Our 10 GHz VLA and FIRST measurements are separated by years to decades, as the FIRST survey includes observations from 1993 to 2011; our measurements are affected by variability too. Our analysis suggests that the quasars in our sample have variability of about a factor of two and a half over the time period considered. This is true for both core-dominated and lobe-dominated quasars. Approximately two fifths of our sample shows excess core flux density in FIRST measurements beyond the amount ascribable to variability. Hence, our results demonstrate statistically that the contamination of FIRST core measurements by the extended emission is apparent, as discussed by \cite{JB2013}.

Also important in the determination of radio core dominance is the choice of the normalization factor.  We have reported radio core dominance measurements by normalizing the radio cores by both the extended radio flux density and the optical. The optical normalization of $\rm R_{5100}$ is preferred. \cite{WBr95} first defined an alternative way to normalize core luminosity by the V-band optical luminosity as a measure of intrinsic jet power. They claimed that $\rm R_{V}$ is a better orientation indicator than R, as it shows a better correlation with the jet angle and FWHM $\rm H{\beta}$ compared to the more traditional $\rm R$. 
While the optical continuum likely has an orientation dependence (e.g., \citealt{NB2010}), it is relatively small compared to the variation in radio cores seen within the beaming angle.  The optical luminosity is also representative of the power of the AGN at the present epoch, while the extended radio emission represents some time averaging of the power of past activity.  The extended radio luminosity is also affected by the jet's interaction with the environment, introducing additional scatter.  \cite{Van2015} presents a comparative study of several different radio-based radio core dominance normalizations, using complementary tests, also finding the optical continuum a superior normalization choice. %Talk about shift from V-band to monochromatic 5100 flux density

\begin{figure}[h!]
    \centerline{\includegraphics[scale=0.6
    ]{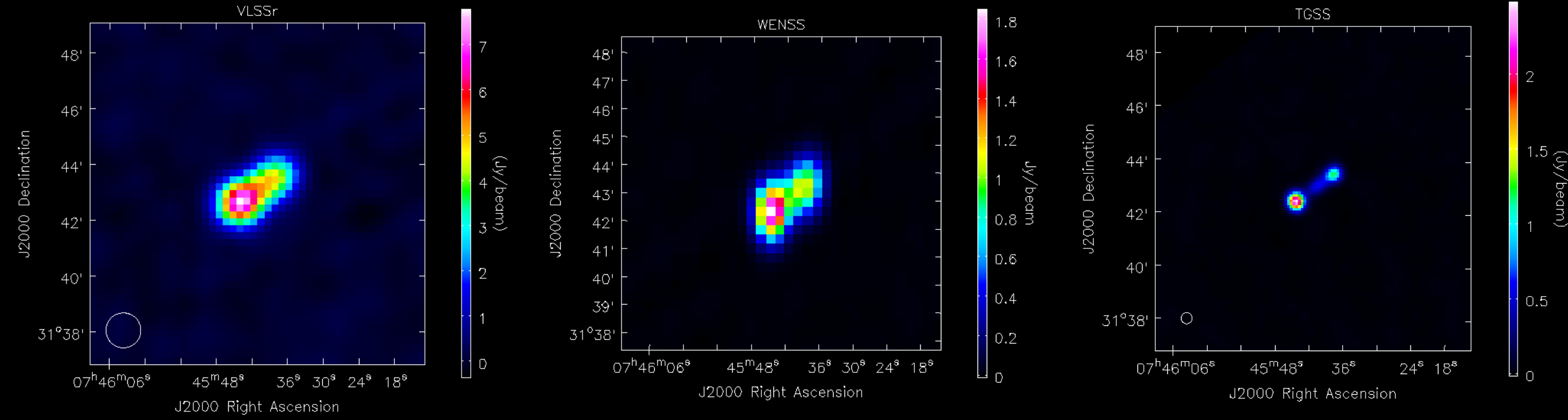}}
    \caption{Radio maps of J074541.66+314256.6 from VLSSr, WENSS and TGSS demonstrating their resolving power.\label{fig:fig7}}
\end{figure}

The extended radio flux density normalization of $\rm R$ is laborious. Low frequency  observations are preferred since extended emission has a steep spectrum compared to the relatively flat spectrum of cores.  We adopted the WENSS total flux density at observed-frame 325 MHz. Even lower frequency observations would be better in principle, as the degree of core contamination will be increasingly negligible, although there are other considerations. \edit1{The lowest frequency sky survey we use is the VLSSr at 74 MHz. It fails to detect fainter sources because of its low sensitivity. Also, at this frequency, the slope of CSS sources shows flattening.} The TGSS at 150 MHz is the highest resolution and sensitivity radio survey available for obtaining the total flux density measurements. It has a resolution of 25$\arcsec$ at which the extended emission may be resolved. \edit1{Figure \ref{fig:fig7} demonstrates the resolving power of these surveys. J074541.66+314256.6 appears as a single source at the resolution of VLSSr and WENSS, but it is a resolved two-component source in TGSS.} Therefore, the search radius to match lobe candidates should be carefully chosen to make sure that all components are collected, otherwise one may end up underestimating the flux densities.  Using lobe matching technique that \cite{JB2020} performed for WENSS, could be very taxing for TGSS as a search radius of 1100$\arcsec$ would result in a large number of detection to match to a core. Another downside with TGSS is that it suffers with a resolution bias. It has a higher resolution power but a poor low surface brightness sensitivity. This leads to lower total flux densities for objects fainter than 2 Jy when compared to a lower resolution survey like 7C \citep{Intema2017}. Although, for our sample we found this threshold to be $\sim 580$ mJy by comparing with 7C survey. Plotting SEDs with total flux densities from other low resolution surveys like WENSS and NVSS shows that 7C measurements comply with the overall trend, whereas TGSS measurements for the fainter objects fall below because of the resolution bias. For a small set of objects, TGSS is the preferred survey for extended flux densities given that the objects are bright enough and the matching is done carefully. For a large set of objects, this would be difficult, in which case WENSS falls at the sweet spot of frequency and resolution good for extended flux density measurements. Several orientation studies used NVSS extended flux densities as it is at the same frequency as the FIRST. Being at 1.4 GHz, the NVSS total flux density is often beamed and suffers from core variability, hence is not a good choice for normalization.

The extended flux densities needs to be k-corrected (\citealt{Hogg}) and scaled to 5 GHz for the measurement of R. When using low frequency data like from TGSS and WENSS, $\alpha_{ext}$ should be choosen carefully. As the frequency difference is very large, a small change in $\alpha_{ext}$ can significantly affect your R measurements. We took the approach of determining the extended slope for individual objects by fitting their SEDs from TGSS, 7C, WENSS, and NVSS (Appendix \ref{sec: app1}).  In lobe-dominated quasars this gives a fair measurement of the radio spectral index for extended emission and we can account for variability concerns separately.  For core-dominated sources, $\alpha_{ext}$  is not well measured in this way and requires simultaneous measurements. So, we adopted the mean spectral index of our sample for them.

The actual utility of radio core dominance is as an indicator of the orientation of the radio jet to the line of sight, and it is desirable to be able to relate it to a physical angle. \cite{MarinA2016} provide such a quantitative prescription, deriving a semi-empirical function between core-dominance and inclination angle by numerically fitting a polynomial regression relation to sources within a equally distributed solid angle. Their relation provides a measurement of inclination from log R of radio-loud AGNs with an accuracy of $\approx 10 ^{\degree}$ or less. AGN inclination angle ($i$) can be obtained from the log R (LR) at 5 GHz using
\begin{equation}
     i = g + h(\mbox{LR}) + j(\mbox{LR})^2 + k(\mbox{LR})^3 + l(\mbox{LR})^4 + m(\mbox{LR})^5,
\end{equation}
where, $g = 41.799\pm1.190$, $h = -20.002\pm1.429$, $j = -4.603\pm1.347$, $k = 0.706\pm0.608$, $l = 0.633\pm0.226$ and $m = 0.062\pm0.075$ (see section 3 of \citealt{MarinA2016} for details).

\edit1{The assumption of radio core dominance may be used as a statistical orientation indicator  is based in part on the notion that the structure of the source is constant for a very long time. The jet axis of AGN is known to exhibit precession and can lead to an irregular radio morphology in a fraction of objects experience significant torques.  The time period of jet precession in binary SMBHs is in range $10^3$ - $10^8$ years \citep{Gower,Dunn}. In a study of X-shaped radio galaxies, 21$\%$ are known to be genuinely formed by mergers and only 1.3$\%$ are formed by the spin-flip mechanism \citep{Roberts}. There are few cases in which jet-precession in seen in quasars. QSO 2300$-$189 is one such example where an S-shaped radio morphology is attributed to jet-precession due to a merger with a nearby galaxy \citep{Hunstead}. The bending of jets at kpc scale may not be from time-dependent precession but rather an environmental change such as a density gradient.  Precession, environment, or episodic nuclear activity can all potentially contribute to unusual behavior in individual objects and create outliers in plots involving radio core dominance against other parameters that can be orientation dependent.}

There are several immediate applications of our work to recent investigations. For instance, \cite{Brotherton2015} use a sample of radio-loud quasars selected using the FIRST survey to examine and offer corrections to orientation-biases for black hole mass estimates, as well as propose a radio-quiet quasar orientation indicator.  Because of the high-frequency of FIRST, their sample is deficient in lobe-dominant quasars and biased toward flat spectrum sources. Moreover, their radio core dominance values are from FIRST and have contaminated cores. A similar investigation based on our sample and the new VLA observations would be superior. \cite{JB2020} use a subset of this sample to investigate the dependence of quasar optical spectral properties on orientation.  They focus on updating the \cite{WB86} plot of H$\beta$ FWHM against radio core dominance, finding that the sharp envelope of previous studies where only edge-on sources display the broadest lines is absent. Again, they use FIRST for their radio core measurements and a revised study would benefit from our 10 GHz observations to determine core flux densities. There are many additional applications, for instance, such as the investigation of the association of orientation with quasar spectral principal components (e.g., \citealt{Ma2019}).  Such follow-ups will be the subject of a future paper.

\section{Conclusions} \label{sec:conclusion}
In this paper, we have used a relatively unbiased sample designed for studying orientation effects in radio-loud quasars. The sample is based on matching $0.1 < z < 0.6$ SDSS quasars with the WENSS survey at 325 MHz with a total radio luminosity cutoff $\rm log(L_{325}) > 33.0 ~erg ~s^{-1} ~Hz^{-1}$. At this low-frequency the isotropic extended emission dominates which enables selecting representative quasars relatively unbiased by orientation. In the case of quasar cores, we expect a flat radio-spectrum over a broad range of frequencies. Jackson \& Browne (2013) pointed out that most survey measurements of core flux densities, like FIRST, often do not have the spatial resolution to always distinguish cores from the extended regions. We present observational evidence for this claim by comparing FIRST measurements of core flux densities with newly obtained 10 GHz VLA A-array measurements.  We discern a systematically higher core brightness in the case of FIRST measurements for about two fifths of our sample, even after taking into account core variability (Figure \ref{fig:fig4}). We present examples in Appendix \ref{sec: app2} showing resolved structure in the 10 GHz image as compared to point-like structure in FIRST image.

We compared the two core dominance parameters R and $\rm R_{5100}$ using core flux densities from the FIRST and the new observations (Figure \ref{fig:fig4}). Use of FIRST measurements resulted in a systematically higher value of core dominance, hence more quasars appeared core-dominated. Overall we recommend using our optically normalized core dominance parameter $\rm R_{5100, 10 GHz}$ based on other studies in the literature.

Our results provide improved measurements of the core flux densities and the core dominance parameters that have potential implication in understanding accurate correlation between radio and optical properties of the quasars. This sample will also serve to update several past studies that employed FIRST-based core dominance measurements.

\acknowledgments

We would like to thank Todd Boroson for his contribution to the sample selection and Sally Laurent-Muehleisen for her insights in FIRST survey. We thank the anonymous referee for their constructive remarks that helped to imporve the manuscript.

The National Radio Astronomy Observatory is a facility of the National Science Foundation operated under cooperative agreement by Associated Universities, Inc.

Funding for the SDSS and SDSS-II has been provided by the Alfred P. Sloan Foundation, the Participating Institutions, the National Science Foundation, the U.S. Department of Energy, the National Aeronautics and Space Administration, the Japanese Monbukagakusho, the Max Planck Society, and the Higher Education Funding Council for England. The SDSS Web Site is http://www.sdss.org/.

The SDSS is managed by the Astrophysical Research Consortium for the Participating Institutions. The Participating Institutions are the American Museum of Natural History, Astrophysical Institute Potsdam, University of Basel, University of Cambridge, Case Western Reserve University, University of Chicago, Drexel University, Fermilab, the Institute for Advanced Study, the Japan Participation Group, Johns Hopkins University, the Joint Institute for Nuclear Astrophysics, the Kavli Institute for Particle Astrophysics and Cosmology, the Korean Scientist Group, the Chinese Academy of Sciences (LAMOST), Los Alamos National Laboratory, the Max-Planck-Institute for Astronomy (MPIA), the Max-Planck-Institute for Astrophysics (MPA), New Mexico State University, Ohio State University, University of Pittsburgh, University of Portsmouth, Princeton University, the United States Naval Observatory, and the University of Washington.
%% To help institutions obtain information on the effectiveness of their 
%% telescopes the AAS Journals has created a group of keywords for telescope 
%% facilities.
%
%% Following the acknowledgments section, use the following syntax and the
%% \facility{} or \facilities{} macros to list the keywords of facilities used 
%% in the research for the paper.  Each keyword is check against the master 
%% list during copy editing.  Individual instruments can be provided in 
%% parentheses, after the keyword, but they are not verified.

\vspace{5mm}
\facility{VLA}
\software{CASA }

\appendix
\section{Radio spectral energy distributions}\label{sec: app1}
\edit1{We plot the total flux densities vs frequency for VLSSr, TGSS, 7C, WENSS, NVSS and FIRST. The SEDs also includes core flux densities from FIRST and the new 10 GHz measurements. Figure 8(a) shows the SEDs of Group1 and Group2 sources that are likely core-dominated or intermediate quasars. Figure 8(b) shows the SEDs of Group2 sources that are likely lobe-dominated quasars and Group2-flag sources that are likely CSS quasars. The total flux densities from VLSSr, 7C, TGSS, WENSS \& NVSS are fitted with a slope to calculate $\rm \alpha_{ext}$. The error in slope is given by the fitting algorithm as described in the main text.} 

\begin{figure}[p]
\epsscale{1}
\plotone{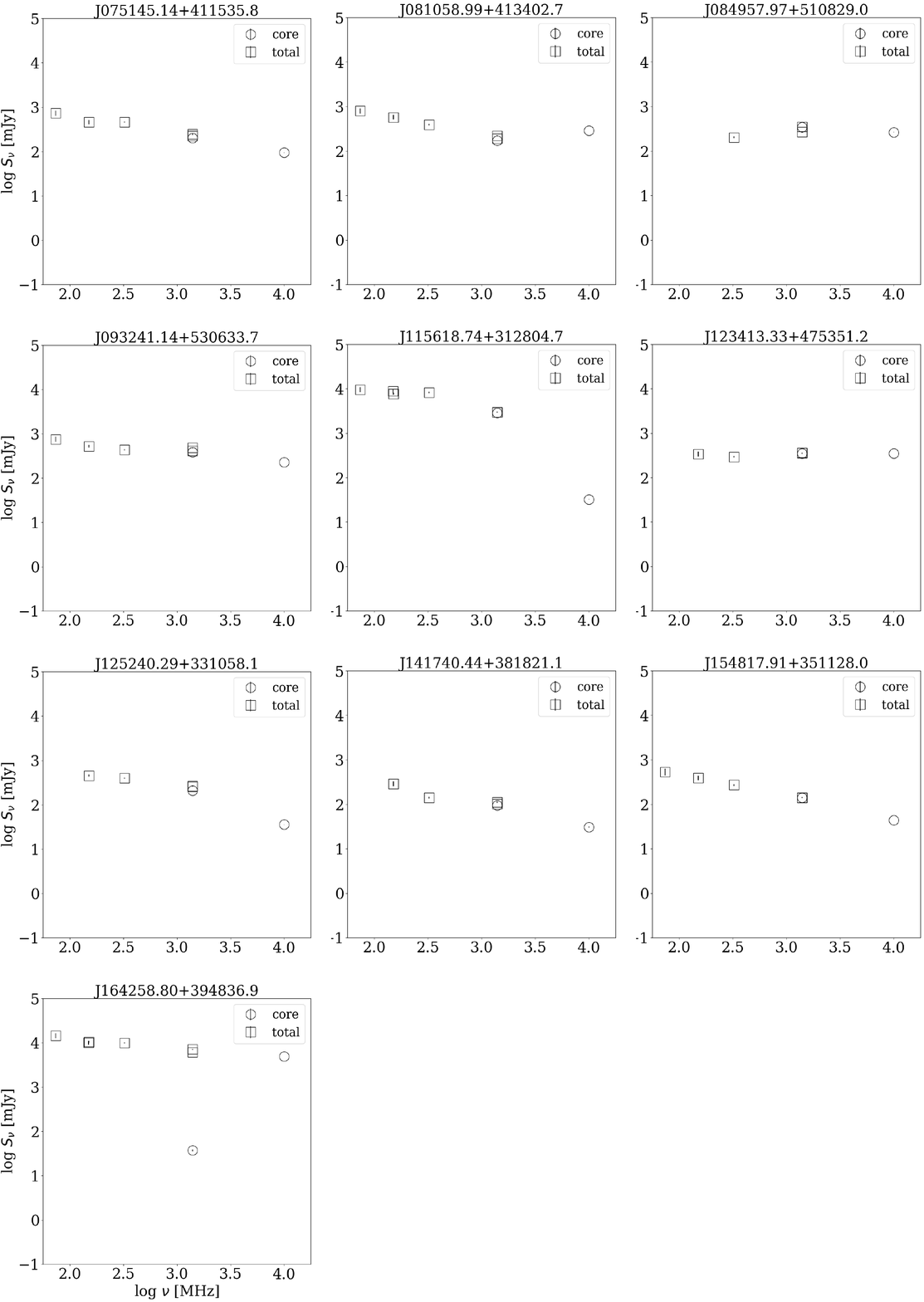}
\caption{\textbf{(a)} Log-log plot of the radio flux density vs frequency. The data points include total flux densities from VLSSr, 7C, TGSS, WENSS, FIRST \& NVSS. It also includes core flux densities from FIRST and the new 10 GHz measurement. Spectral slopes are not plotted for core-dominated and intermediate (Group 1 \& 3) quasars as they would be affected by core variability.}
\end{figure}
\addtocounter{figure}{-1}
\begin{figure}[p]
\epsscale{1}
\plotone{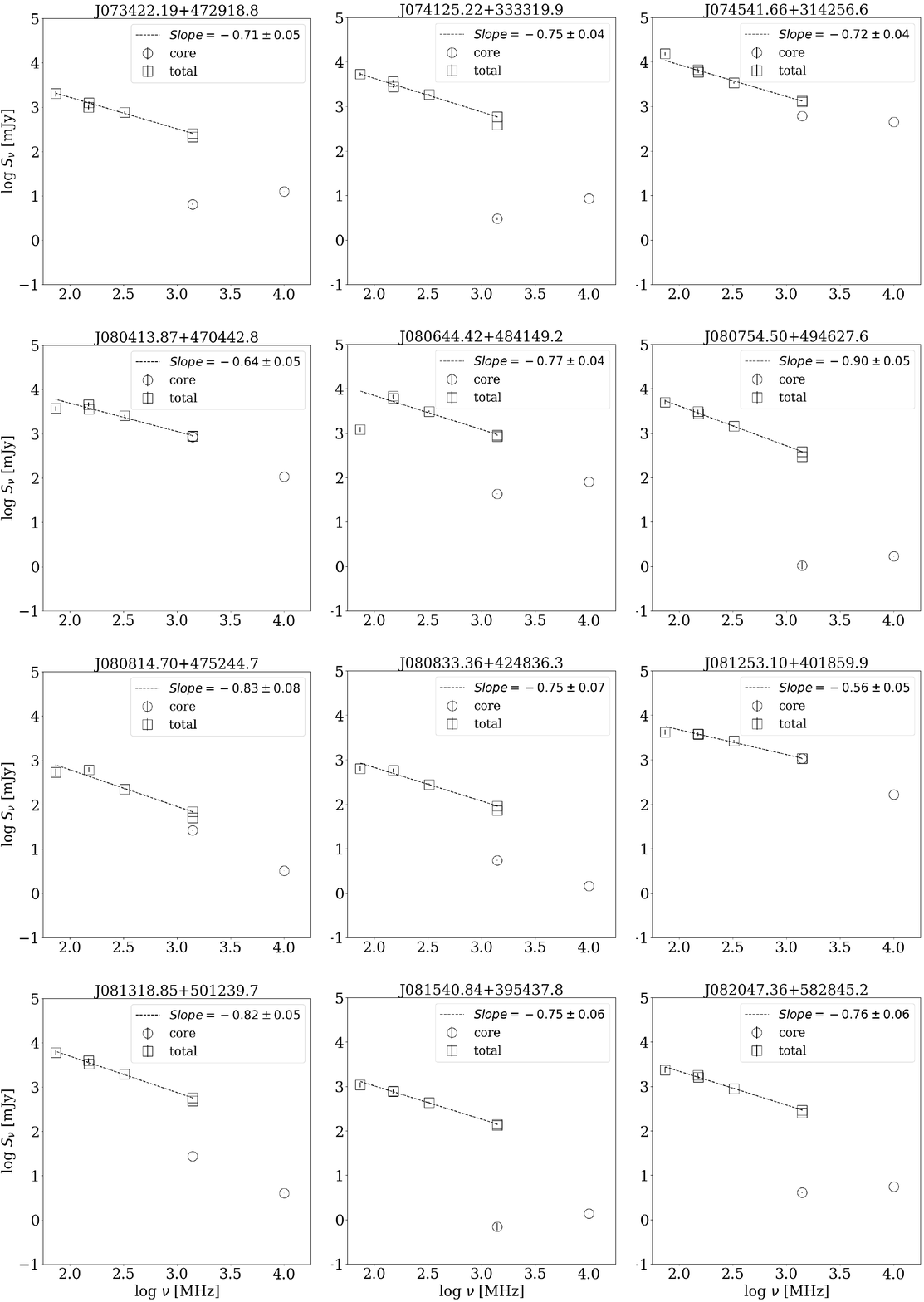}
\caption{\textbf{(b)} Log-log plot of the radio flux density vs frequency. The data points include total flux densities from VLSSr, 7C, TGSS, WENSS, FIRST \& NVSS. It also includes core flux densities from FIRST and the new 10 GHz measurement. Dashed line shows the spectral index of the extended emission obtained by fitting a slope to the total flux densities from VLSSr, 7C, TGSS, WENSS \& NVSS. Spectral slopes are plotted for Group 2 \& Group 2-flag quasars.}
\end{figure}
\renewcommand{\thefigure}{\arabic{figure} (b) Continued}
\addtocounter{figure}{-1}
\begin{figure}[p]
\epsscale{1}
\plotone{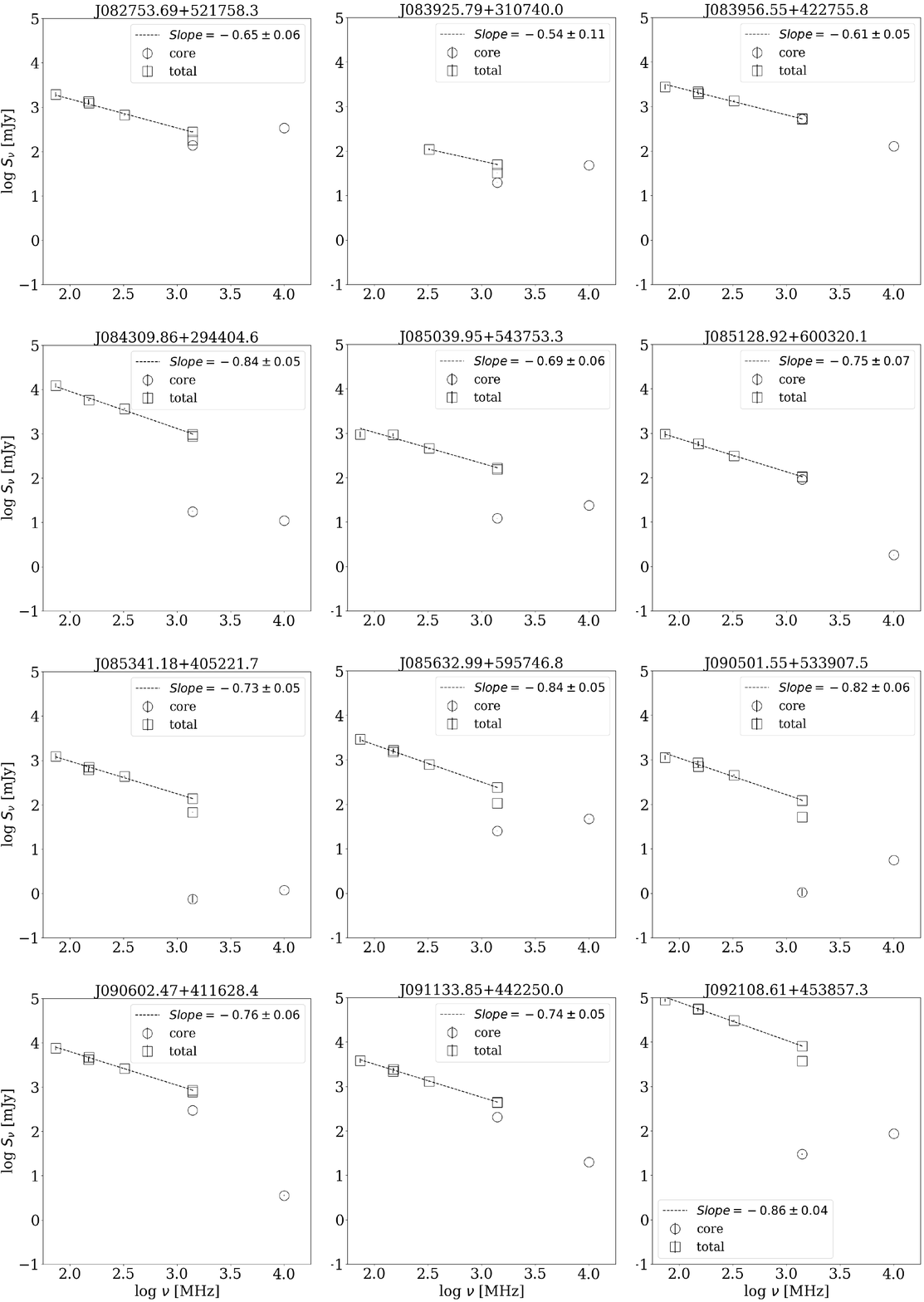}
\caption{}
\end{figure}
\renewcommand{\thefigure}{\arabic{figure} (b) Continued}
\addtocounter{figure}{-1}
\begin{figure}[p]
\epsscale{1}
\plotone{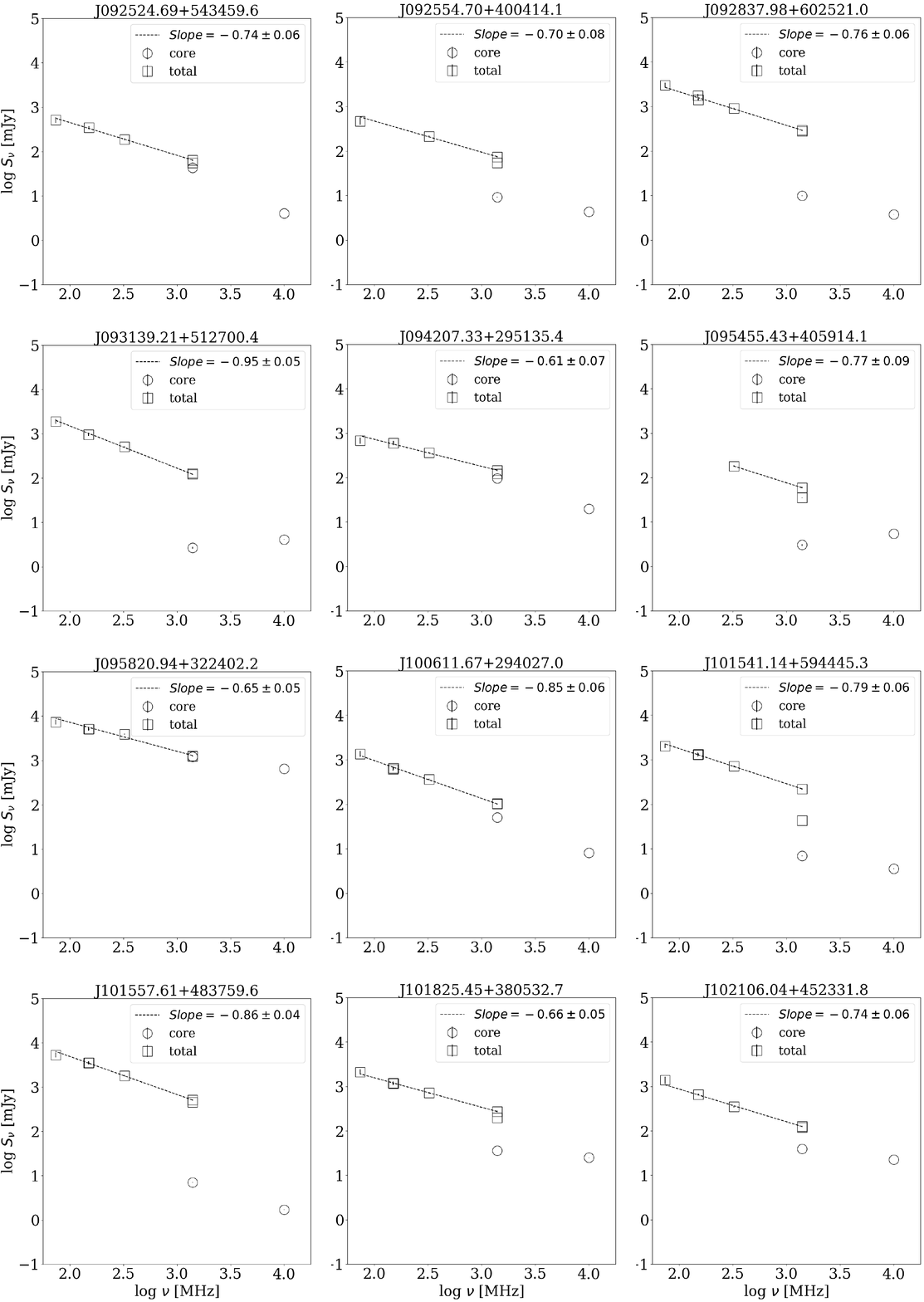}
\caption{}
\end{figure}
\renewcommand{\thefigure}{\arabic{figure} (b) Continued}
\addtocounter{figure}{-1}
\begin{figure}[p]
\epsscale{1}
\plotone{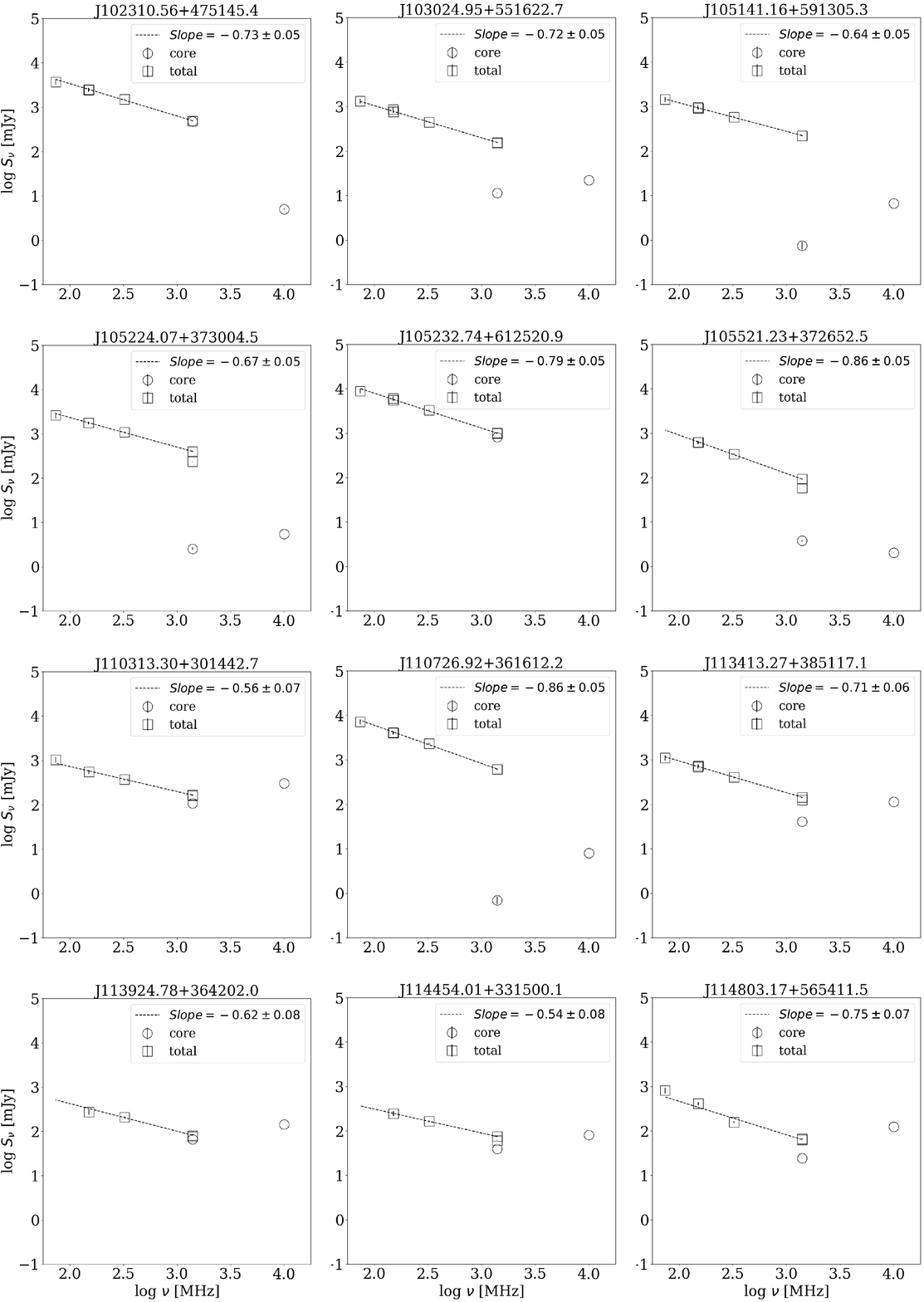}
\caption{}
\end{figure}
\renewcommand{\thefigure}{\arabic{figure} (b) Continued}
\addtocounter{figure}{-1}
\begin{figure}[p]
\epsscale{1}
\plotone{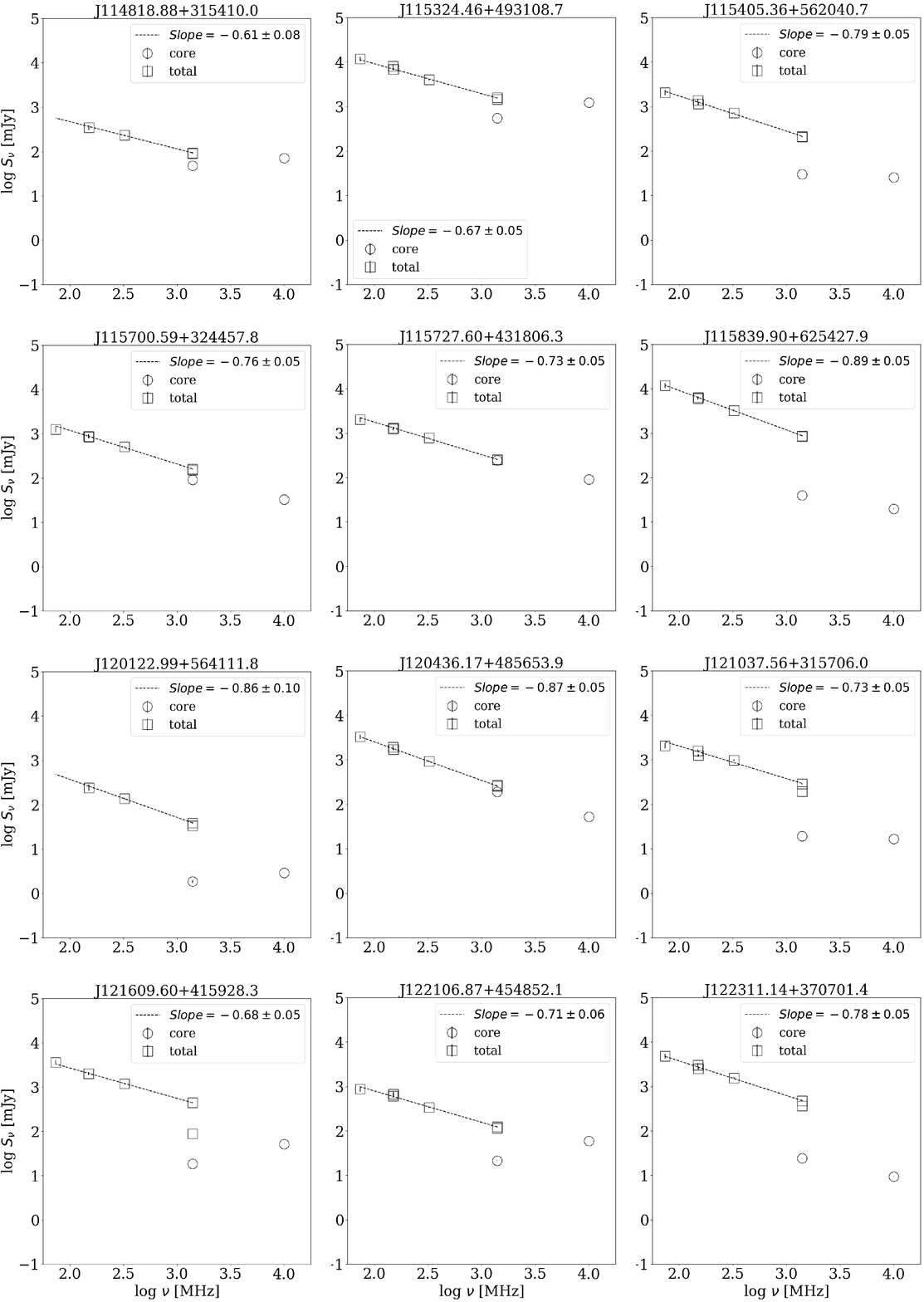}
\caption{}
\end{figure}
\renewcommand{\thefigure}{\arabic{figure} (b) Continued}
\addtocounter{figure}{-1}
\begin{figure}[p]
\epsscale{1}
\plotone{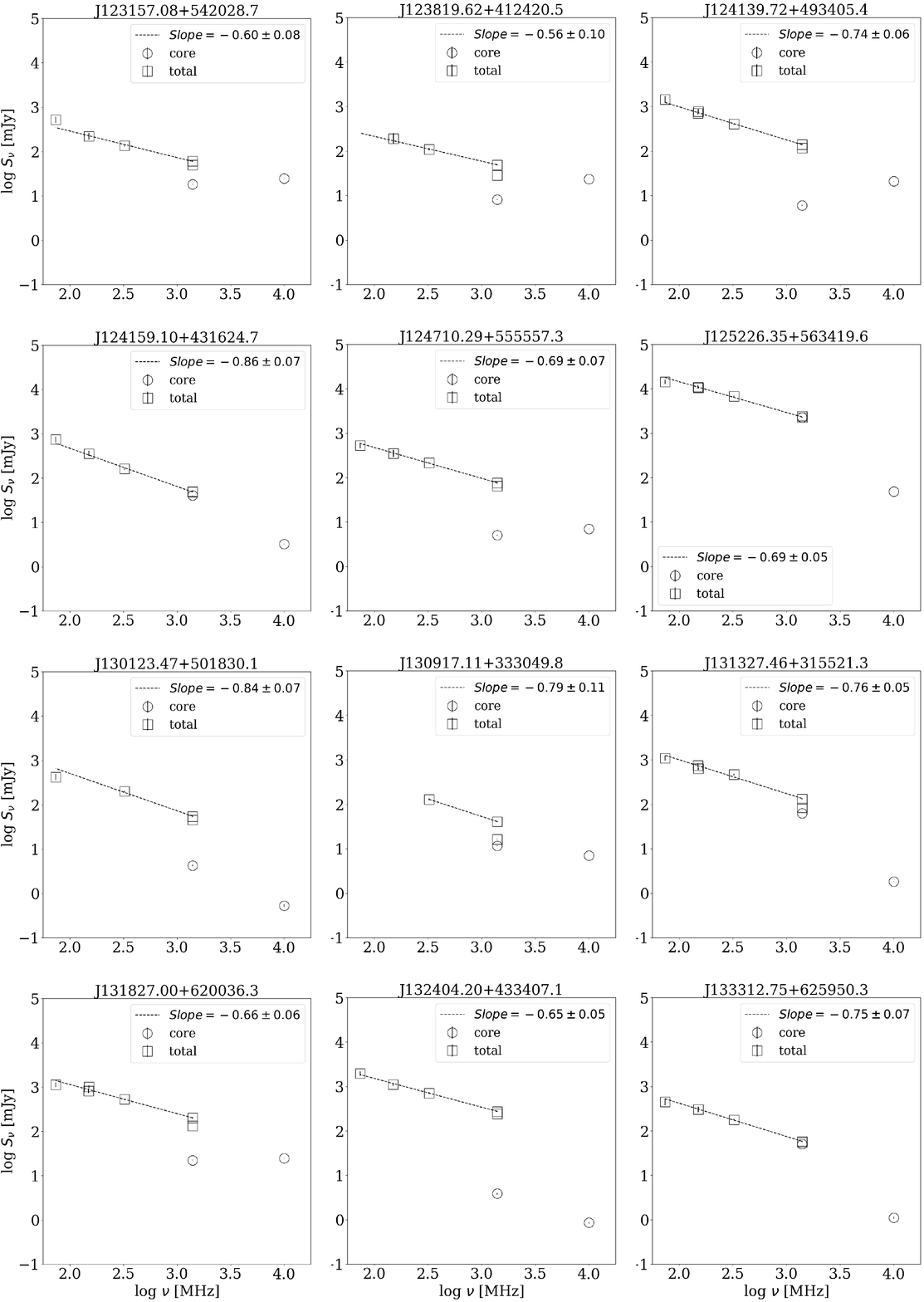}
\caption{}
\end{figure}
\renewcommand{\thefigure}{\arabic{figure} (b) Continued}
\addtocounter{figure}{-1}
\begin{figure}[p]
\epsscale{1}
\plotone{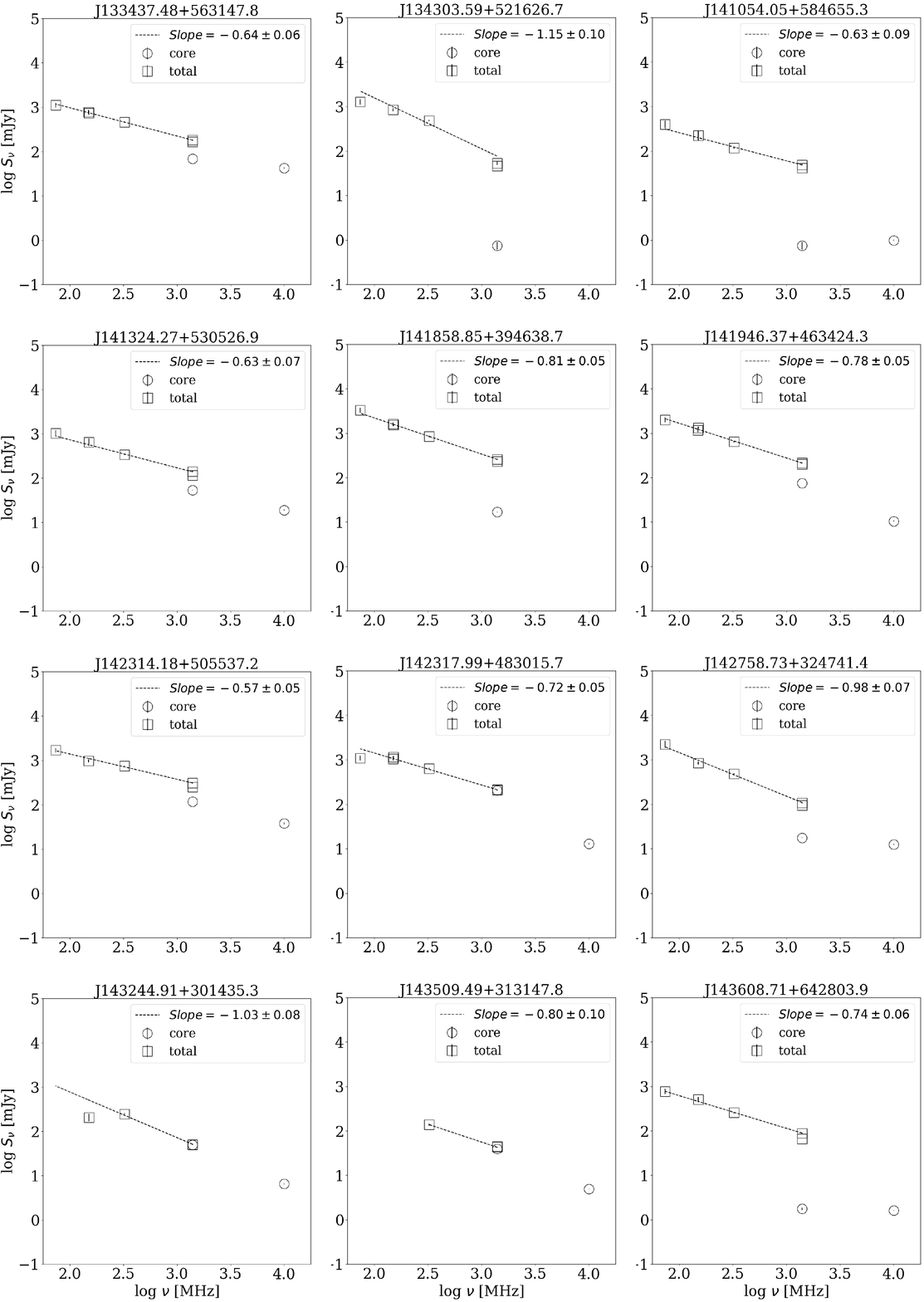}
\caption{}
\end{figure}
\renewcommand{\thefigure}{\arabic{figure} (b) Continued}
\addtocounter{figure}{-1}
\begin{figure}[p]
\epsscale{1}
\plotone{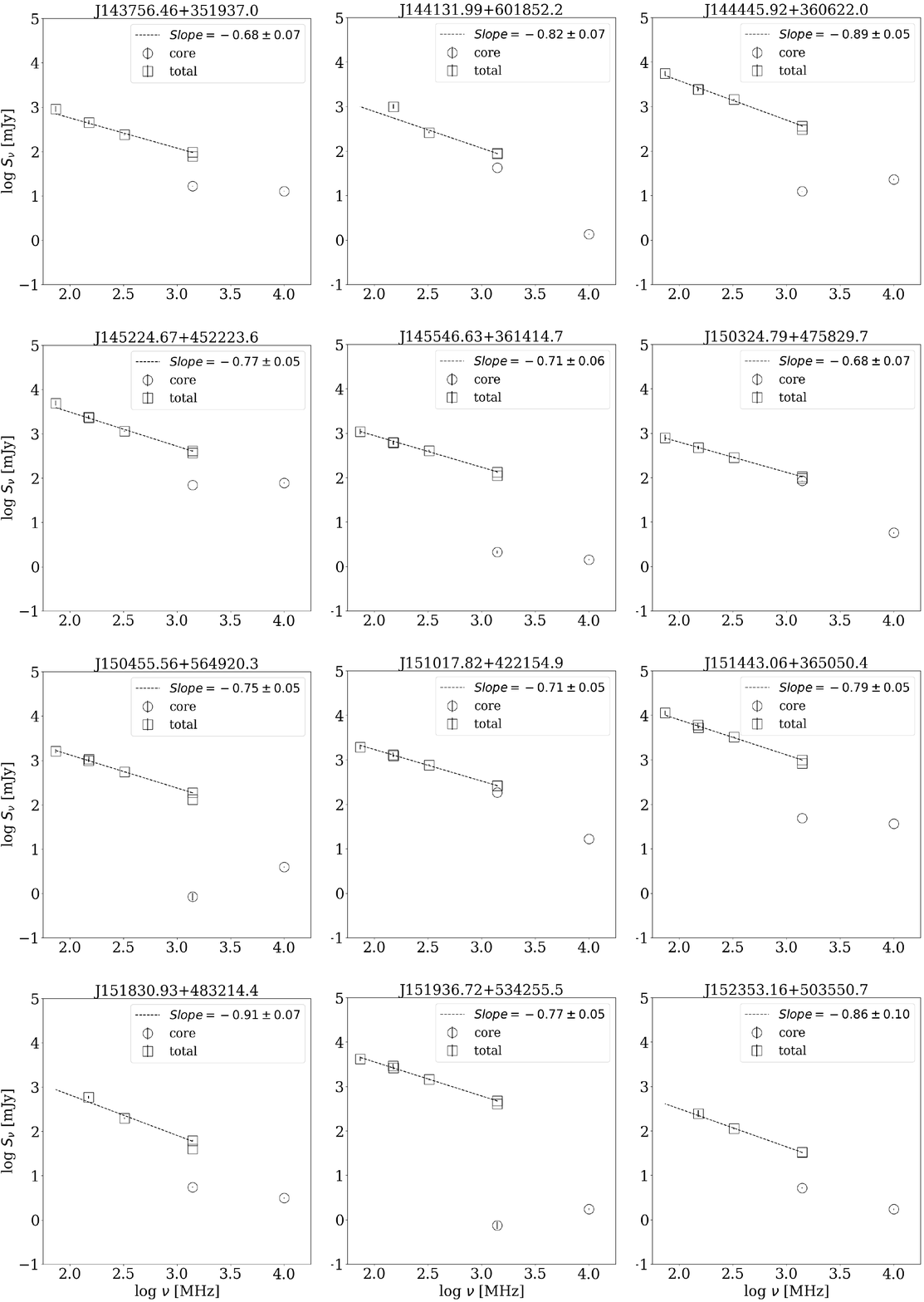}
\caption{}
\end{figure}
\renewcommand{\thefigure}{\arabic{figure} (b) Continued}
\addtocounter{figure}{-1}
\begin{figure}[p]
\epsscale{1}
\plotone{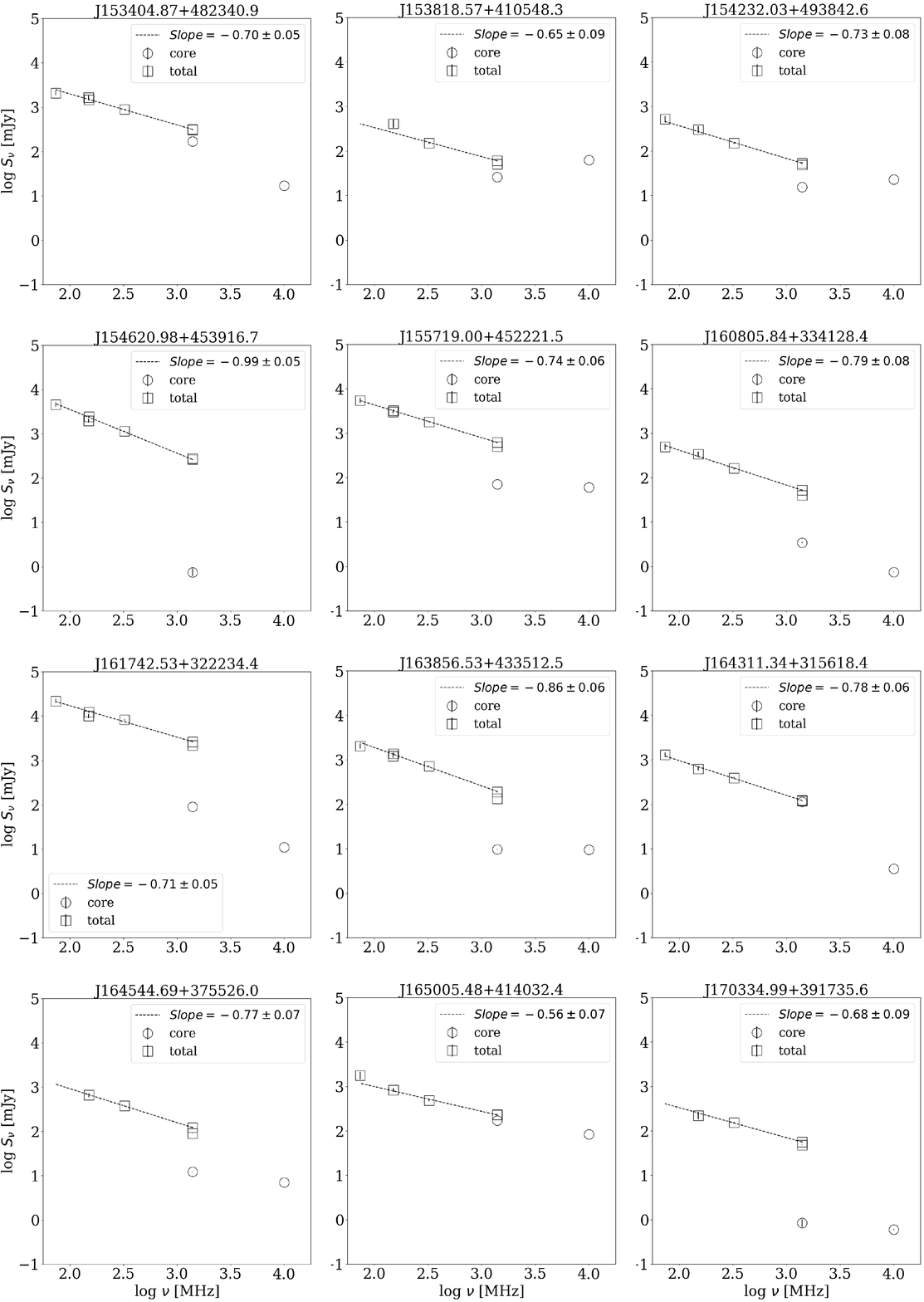}
\caption{}
\end{figure}
\renewcommand{\thefigure}{\arabic{figure} (b) Continued}
\addtocounter{figure}{-1}
\begin{figure}[p]
\epsscale{1}
\plotone{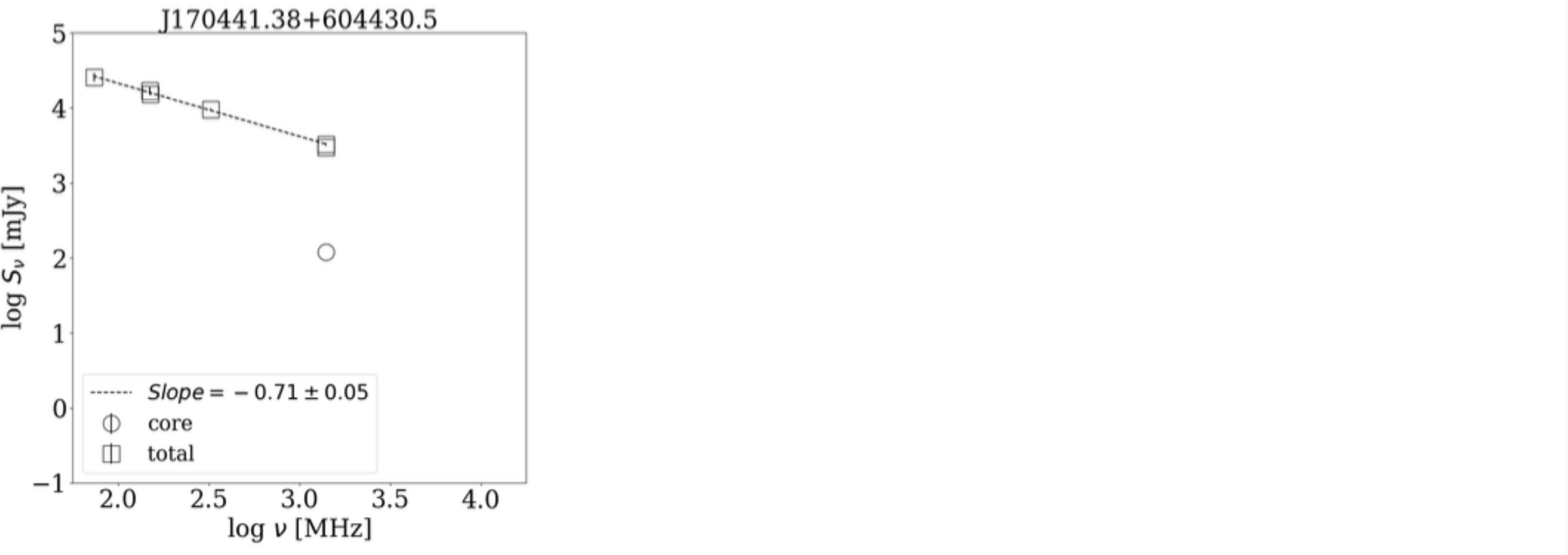}
\caption{}
\end{figure}
\renewcommand{\thefigure}{\arabic{figure}}

\newpage

\section{Sources with resolved 10 GHz images}\label{sec: app2}

\begin{deluxetable}{crrc}[h!]
\tablecaption{ RMS, mean and dynamical range of the resultant maps of VLA 10 GHz presented in this paper. \label{tab:table3}}
\tablehead{
\colhead{Object name} & \colhead{RMS} & \colhead{Mean} & \colhead{Dynamical Range}\\
\colhead{SDSS} & \colhead{Jy beam$^{-1}$} & \colhead{Jy beam$^{-1}$} & \colhead{}
}
\colnumbers
\startdata
J080413.87+470442.8	&	2.44E-03	&	1.15E-04	&	44.05\\
J085128.92+600320.1	&	8.66E-05	&	1.34E-05	&	21.06\\
J100611.67+294027.0	&	1.34E-04	&	5.62E-06	&	60.59\\
J124159.10+431624.7	&	1.13E-04	&	1.24E-05	&	28.89\\
J125226.35+563419.6	&	1.54E-02	&	2.08E-03	&	20.62\\
J164311.34+315618.4	&	2.61E-04	&	5.74E-05	&	13.72\\
J170334.99+391735.6	&	2.72E-05	&	6.08E-07	&	22.13\\
\enddata
\end{deluxetable}
\begin{figure}[h!]
\epsscale{1.1}
\plotone{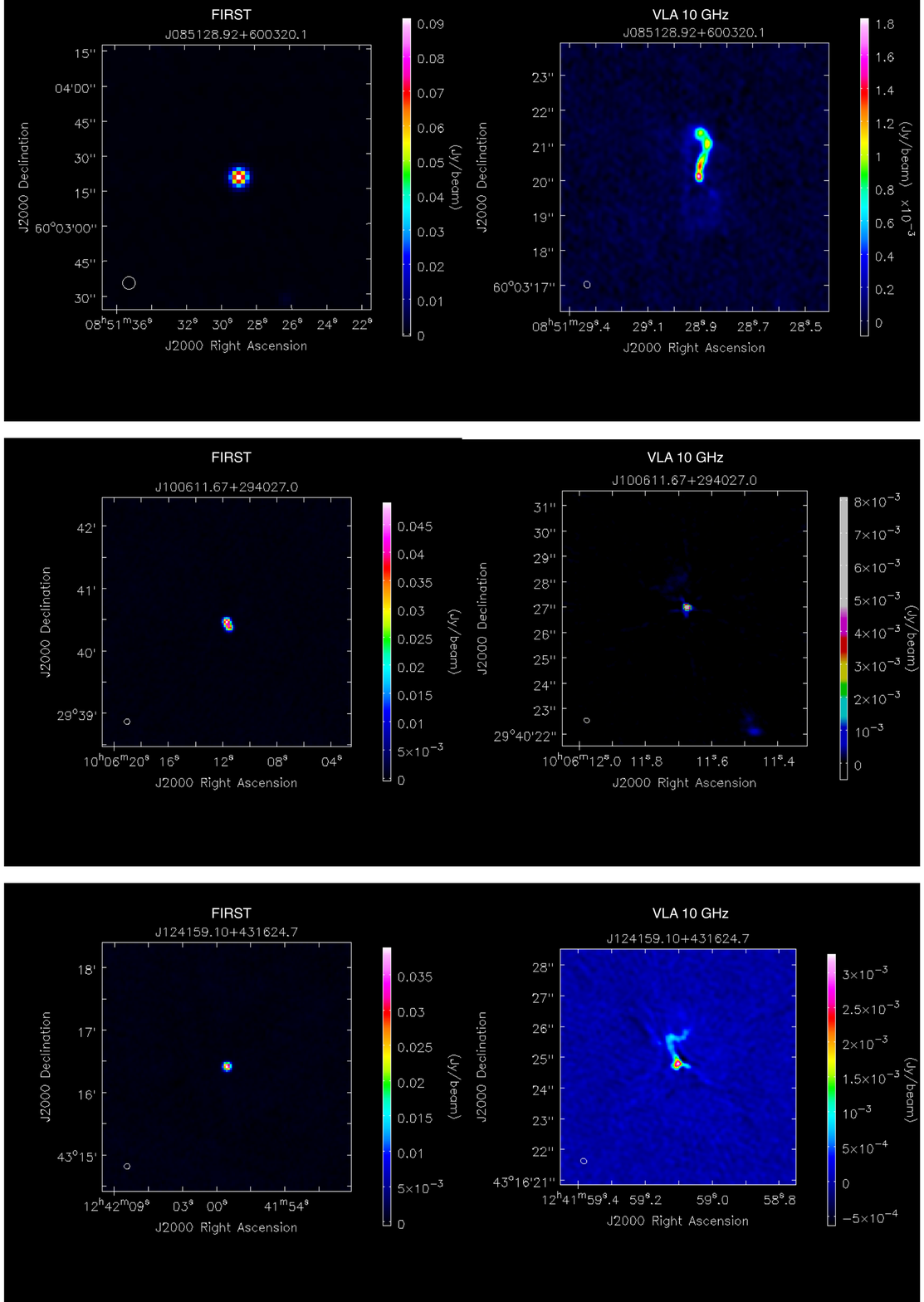}
\caption{The FIRST image (on the left) of half a dozen targets compared to the new observation at 10 GHz (on the right). The high resolution 10 GHz images show extended structures that are unresolved at FIRST resolution.}
\end{figure}
\renewcommand{\thefigure}{\arabic{figure} (Cont.)}
\addtocounter{figure}{-1}
\begin{figure}[p]
\epsscale{1}
\plotone{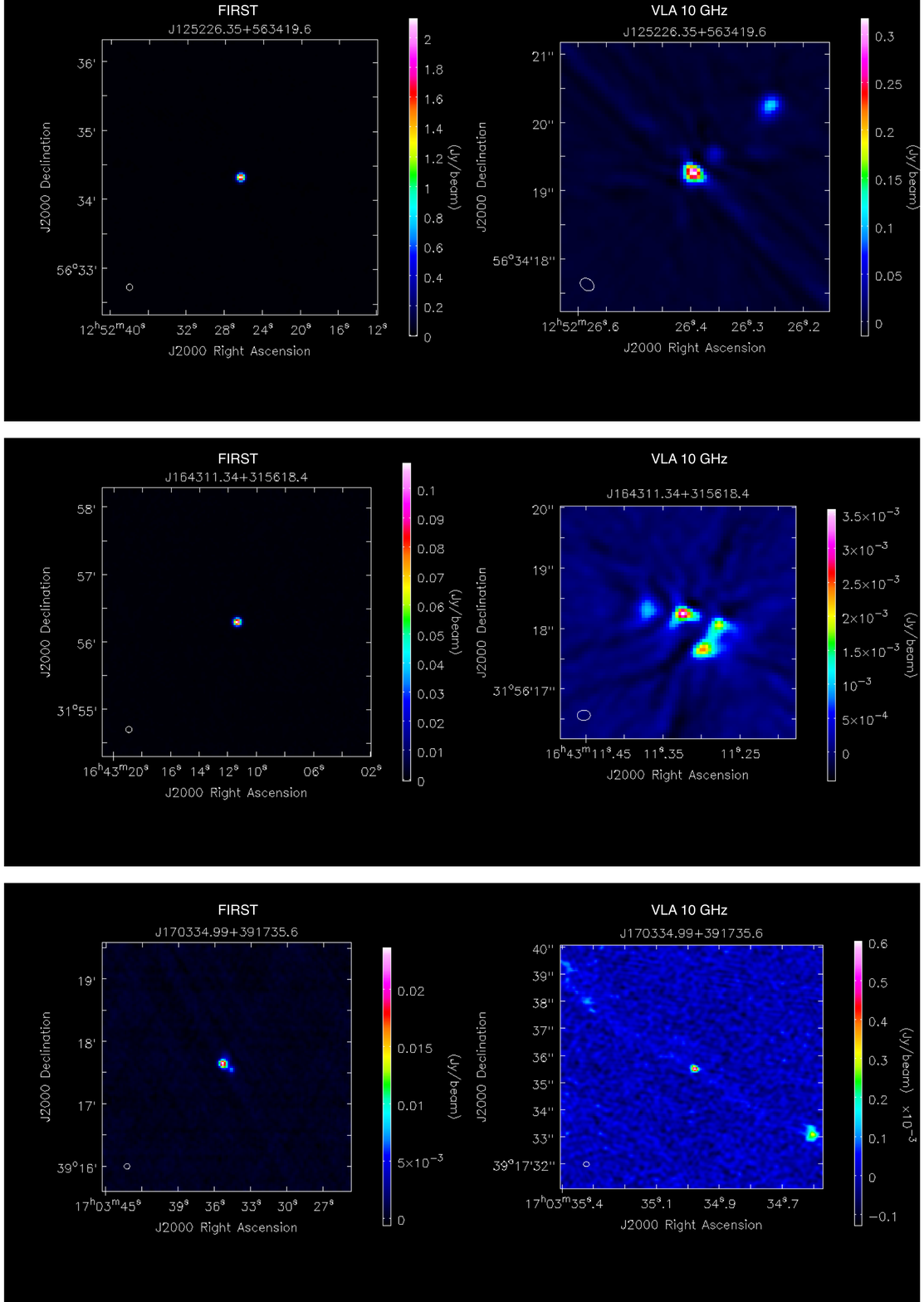}
\caption{}
\end{figure}
\renewcommand{\thefigure}{\arabic{figure}}

\newpage
\section{Additional objects}\label{sec: app0}

\edit1{We proposed VLA observations on a preliminary version of the \citeauthor{JB2020} sample, consisting of 147 radio-loud quasars. They excluded 28 targets after a revision in the luminosity cut. In the spirit of publicizing all available VLA data, we present the measurements in Table 4 available as a machine-readable table. The optical spectra of these targets were analyzed together, thus their spectral properties were calculated identically to the final \citeauthor{JB2020} sample. One of the additional 28 objects has bad 10 GHz data. The remaining 27 were assigned groups based on their SEDs.  Two targets are in Group1, 17 in Group2, five in Group2-flag, and three in Group3.} 

\edit1{We redid the the whole analysis with the additional targets included. Figure 10 compares the core flux densities, R and R5100 measurements. The core variability is a factor of 2.45 for the larger sample. For 53 targets the FIRST core flux density is above the dotted line representing the core variability limit. The means $\alpha_{ext} = -0.746\pm0.130$ is used to k-correct the extended flux density to 5 GHz for Group1 and Group3 targets. The radio core dominance calculated using FIRST measurements suffers from the resolution effect.}

\begin{splitdeluxetable*}{cccccccccBcccccc}
\tablecaption{Data for 28 additional objects observed with VLA.}
\tabletypesize{\scriptsize}
\tablehead{
\colhead{Object name} & \colhead{Redshift} & \colhead{\tablenotemark{$_1$}F5100} & \colhead{$\rm S^{core}_{10 GHz}$} & \colhead{$\rm S^{core}_{1.4 GHz}$} & \colhead{$\rm S_{WENSS}^{total}$} & \colhead{$\rm S_{TGSS}^{total}$} & \colhead{$\rm S_{7C}^{total}$} & \colhead{$\rm S_{VLSSr}^{total}$} & \colhead{$\rm \alpha_{ext}$}  & \colhead{\tablenotemark{$_2$}Group} &
\colhead{log($\rm R_{1.4 GHz}$)} & \colhead{log($\rm R_{10 GHz}$)} & \colhead{log($\rm R_{5100, 1.4 GHz}$)} & \colhead{log($\rm R_{5100, 10 GHz}$)} \\
\colhead{SDSS} & \colhead{z} & \colhead{} & \colhead{$(\rm mJy~beam^{-1})$} & \colhead{$(\rm mJy~beam^{-1})$} & \colhead{$(\rm mJy$)} & \colhead{$(\rm mJy$)} & \colhead{$(\rm mJy$)}  & \colhead{$(\rm mJy$)} & \colhead{} & \colhead{} & \colhead{} & \colhead{} & \colhead{} & \colhead{} 
}
\colnumbers
\startdata
J081432.11+560956.6 & 0.5093 & 7.23$\pm$0.25 & 19.7$\pm$0.07 & 69.18$\pm$0.16 & 67$\pm$6 & 124.2$\pm$13.2 & \nodata & \nodata & -0.746$\pm$0.130 & g3 & 1.08$\pm$0.053 & 0.53$\pm$0.053 & 3.04$\pm$0.03 & 2.50$\pm$0.03 \\
J084925.89+564132.1 & 0.5882 & 5.71$\pm$0.27 & 1.02$\pm$0.02 & 1.42$\pm$0.15 & 62$\pm$7 & 119.5$\pm$12.1 & \nodata & \nodata & -0.688$\pm$0.194 & g2 & -0.95$\pm$0.226 & -1.10$\pm$0.201 & 1.46$\pm$0.12 & 1.31$\pm$0.05 \\
\enddata
\tablenotetext{1}{F5100 is in units of $\rm 10^{-17} erg~ s^{-1} \AA^{-1} cm^{-2}$.}
\tablenotetext{2}{Quasars are assigned group 1, 2, 2-flag (2f) or 3 on the basis of their radio spectrum. See section 5 for details.}
\tablecomments{Table 4 is published in its entirety in the machine readable format.  A portion is shown here for guidance regarding its form and content.}
\end{splitdeluxetable*}

\begin{figure*}[p]
\epsscale{0.83}
\plotone{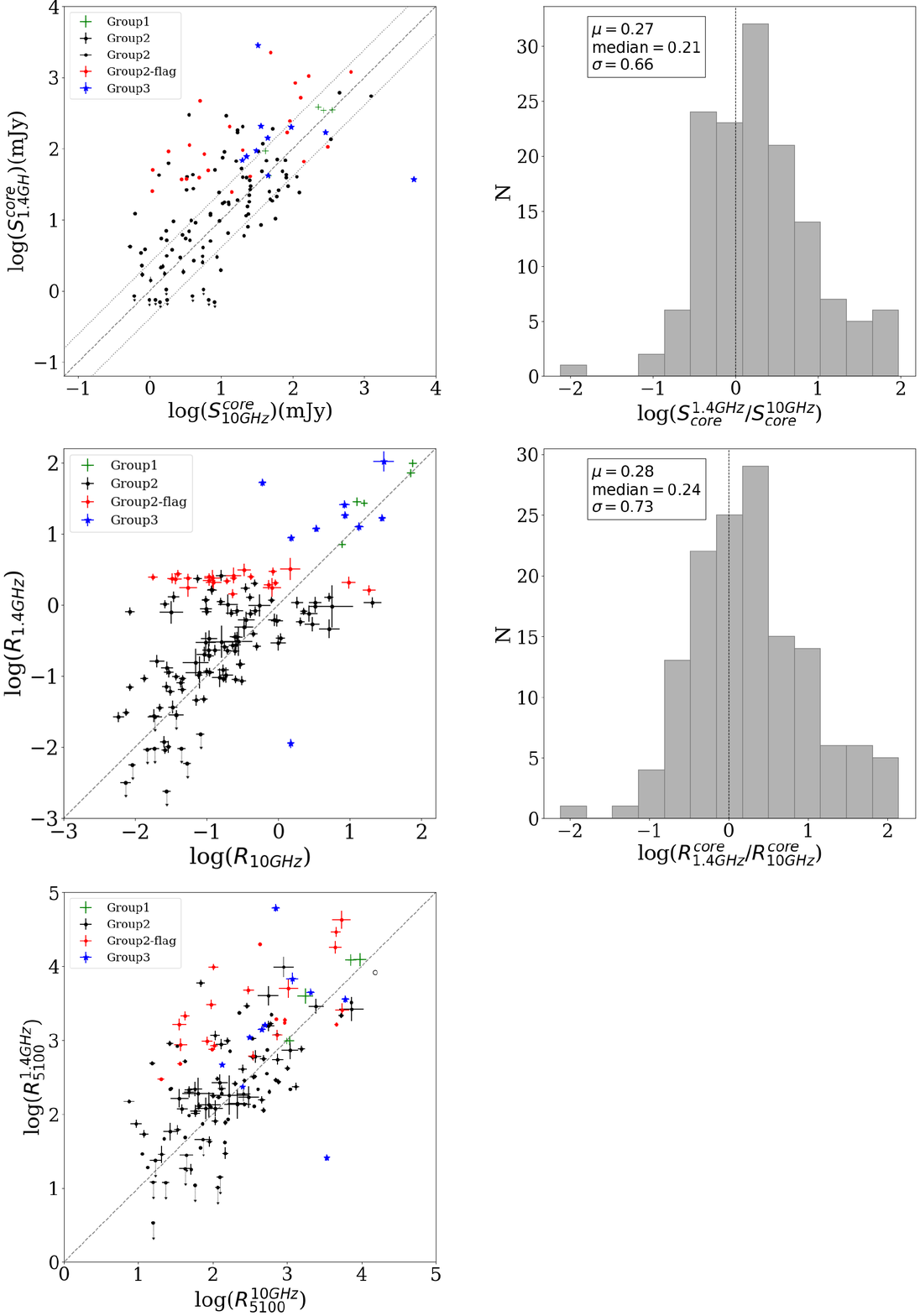}
\caption{The left panels show scatter plots of core flux density measured at 1.4 GHz and 10 GHz and the corresponding R and $\rm R_{5100}$ measurements on log scales. The points are color coded according to the groups defined by the radio spectra of their extended emission. Points on the bottom left corner panel with downward arrows represent the objects with FIRST flux limits. The dashed lines show where the two quantities are equal. \textbf{Top left:} A scatter plot with the logarithm of 1.4 GHz core flux density on y-axis and the logarithm of 10 GHz core flux density on the x-axis.  The dotted line represents $\rm y= x \pm 0.39$, an estimate of the $1\sigma$ scatter introduced by the core variability. \textbf{Middle left:} scatter plot of log R calculated from the 1.4 GHz FIRST cores on the y-axis and from the 10 GHz cores on the x-axis normalized by the extended flux densities scaled to 5 GHz rest-frame according to the groups as discussed in section \ref{sec:R and R5100}. \textbf{Bottom left:} scatter plot of $\rm R_{5100}$ values calculated from the ratio of 10 GHz vs 1.4 GHz core flux densities and 5100$\mbox{\AA}$ flux density. The point with black open circle represent a quasar with a bad 5100 \AA\ flux density measurement.
{\bf Right:} panels show the histograms of the corresponding ratio of the quantities in the x and y axis of the left panels. The dashed lines in these histograms show where the ratios are equal. \label{fig:fig10}}
\end{figure*}
\newpage
%\bibliography{paper}
\bibliography{R2-sample63}{}
\bibliographystyle{aasjournal}
\listofchanges
\end{document}